# Pore-scale Simulation of Shear-Thinning Fluid Flow using Lattice Boltzmann Method

Jithin M.[1], Nimish Kumar[1], Ashoke De[2], Malay K. Das[3]

Energy Conversion and Storage Laboratory

Indian Institute of Technology Kanpur, Kanpur, UP 208016, India

## Abstract

Present work attempts to identify the roles of flow- and geometric-variables on the *scaling factor* which is a necessary parameter for modeling the apparent viscosity of non-Newtonian fluid in porous media. While idealizing the porous media microstructure as arrays of circular and square cylinders, present study uses multi-relaxation time lattice Boltzmann method to conduct pore-scale simulation of shear thinning non-Newtonian fluid flow. Variation in the size and inclusion ratio of the solid cylinders generates wide range of porous media with varying porosity and permeability. Present study also used stochastic reconstruction technique to generate realistic, random porous microstructures. For each case, pore-scale fluid flow simulation enables the calculation of equivalent viscosity based on the computed shear rate within the pores. It is observed that the scaling factor has strong dependence on porosity, permeability, tortuosity and the percolation threshold, while approaching the maximum value at the percolation threshold porosity. Present investigation quantifies and proposes meaningful correlations between the scaling factor and the macroscopic properties of the porous media.

---

[1] Research Assistant, Department of Mechanical Engineering
[2] Associate Professor, Department of Aerospace Engineering
[3] Associate Professor, Department of Mechanical Engineering, corresponding author, Mail: mkdas@iitk.ac.in, Ph: +915122597359

# 1 Introduction

Flow of non-Newtonian fluids in porous media is of major interest to scientists and engineers due to its wide range of applications in the fields of biology, earth sciences, and engineering (Shabro et al. 2012). One important engineering applications of non-Newtonian fluid flow in porous media is related to the enhancement of oil recovery in petroleum industry. Enhancement in the oil recovery may be achieved, by injecting water mixed with polymeric fluids resulting in non-Newtonian rheology (Lopez et al. 2003; Sorbie et al. 1989; Tosco et al. 2013). Also, a number of processes dealt by chemical and environmental engineers, including deep filtration in fixed bed reactors and distillation towers, biomedical separation towers and environmental cleanup technologies involve the use of non-Newtonian fluids (Tosco et al. 2013). Study of non-Newtonian flow of blood in blood vessels is very important in applications of bio-medical applications including the treatment of cardiovascular diseases such as aneurysms and atherosclerosis (Bernsdorf and Wang 2009; Boyd et al. 2007). One method of treatment of aneurysm involves the insertion of porous material in the bulge of the blood vessel. The effects of such treatment are often studied via computer simulation of flow field inside the aneurysm which requires consideration of non-Newtonian flow of blood in the porous medium.

Simulations of fluid flow in porous media in macroscale, generally uses equations averaged over a Representative Elementary Volume (REV) in which all complexities and details associated with the porous microstructure are absorbed into macroscopic variables such as porosity and permeability. Macroscopic models use empirical and semi-empirical relations, such as Darcy Equation, Blake-Kozeny-Carman relation and Ergun Equation. These Equations correlate the superficial velocity and pressure drop with the properties of porous media such as porosity, permeability and particle diameter. All these empirical relations, however, presupposes Newtonian fluid flow, where fluid viscosity does not vary with shear rate. Non-Newtonian fluids, on the other hand, have variable viscosity which is a function of local shear rate. Different empirical models, like the power-law, Cross, Ellis and the Carreau models, have been proposed to relate the local viscosity of the fluid and the local shear rate. The non-linear relationship between the stress and the rate of strain makes it impossible to model non-Newtonian flow using the correlations of Newtonian fluids.

There have been attempts to model non-Newtonian fluid flow at very low flow rates using simple modification of the Newtonian correlations by the definition of an effective viscosity, called porous medium viscosity, to replace the Newtonian viscosity (Tosco et al. 2013). The porous medium viscosity is related to a porous medium shear rate which indicates the average shear rate experienced by the fluid as it flows through the pores. This shear rate is defined as the ratio of flow rate and the porous medium characteristic length scale, as follows:

$$\dot{\gamma}_m = \alpha \frac{q}{\sqrt{K\varepsilon}} \tag{1}$$

Here, $\dot{\gamma}_m$ is the porous medium shear rate, $q$ is the flow rate, $K$ is the permeability and $\varepsilon$ is the porosity of the medium. $\alpha$ is a scaling factor introduced to match the porous medium equivalent viscosity with the actual viscosity of the fluid. Studies show that while the scaling factor depends on properties of the porous medium and the rheology of fluid (Cannella et al.

1988) the value of $\alpha$ generally lies between 1 and 15 (Lopez et al. 2003; Sorbie et al. 1989; Tosco et al. 2013). A major difficulty associated with modeling non-Newtonian fluid flow is that, in addition to the properties of fluid and porous medium, the knowledge of the scaling factor is required which is not an intrinsic property of the porous medium or the flowing fluid.

The value of scaling factor for each fluid-porous medium pair is to be obtained from experiments or pore-scale simulations. So far, very few attempts have been made to obtain information on scaling factor from pore scale simulations. Sorbie et al. (1989) conducted pore network modeling using 2D capillary array and studied the macroscopic behavior based on the microscopic flow in the pores. They predicted the value of $\alpha$ to be dependent on the pore structure and suggested that the more heterogeneous the pore networks are, the larger can the value be. For power law fluids, they derived a correlation for scaling factor in terms of power law index. Perrin et al. (2006) conducted pore scale simulations by assuming capillary networks and compared values of $\alpha$ obtained with experimentally obtained ones. Another work by Lopez et al. (2003) also used network modeling to study the relationship between scaling factor and flow rate. Pearson and Tardy (2002) suggested that scaling factors will also be affected by tortuosity of the geometry. Studies to find the relation between the scaling factor and the porous medium, presently available in literature, are highly inadequate as most of them idealize the porous medium as capillary bundle or use the pore network modeling. Pore network models or models using averaged equations (Hayes et al. 1996; Liu and Masliyah 1999) cannot effectively conduct such a study as the local viscosity depends on local shear rate, and thus the scaling factor cannot be calculated without detailed representation of pore structure (Sochi 2010).

In the present work, we attempt simulation of non-Newtonian fluid flow in the pore scale to study how the pore geometry of the porous media affects the scaling factor. Initially, regular arrangements of circular or square solids are considered in the reconstruction of the microstructure. The size and arrangement of the bodies are varied to obtain different values of the properties of porous medium like porosity and permeability. The effects of these properties on the scaling factor are quantified after conducting a number of numerical experiments. Apart from the regular arrangements of circular and square solids, simulations are also conducted in stochastically reconstructed realistic porous microstructure. For generating the realistic microstructure, present study uses the Quartet Structure Generation Set (QSGS) technique(Wang et al. 2007; Wang et al. 2006). The theory and application of the QSGS techniques is discussed elsewhere(Wang and Pan 2008; Wang and Pan 2009).

Present study simulates flow of blood, which is a shear thinning fluid, (Psihogios et al. 2007; Rakotomalala et al. 1996), through regular as well as stochastically reconstructed porous media. Many rheological models have been proposed in literature to capture the shear thinning behavior of fluids (Sochi 2010). Power law model is one of the widely used models (Aharonov and Rothman 1993; Boek et al. 2003; Gabbanelli et al. 2005; Kehrwald 2005; Psihogios et al. 2007; Rakotomalala et al. 1996; Sullivan et al. 2006), because of its simple form and availability of analytical solution for many physical problems. However, power law model has a limitation that the viscosity values go to infinity and zero for very low and very high shear rate values, respectively, for shear thinning fluids. Other models, such as Casson model and Carreau-Yasuda model, have been proposed which overcome the problems faced by power law model, as it has inherent constant viscosity at low and high shear rate (Ashrafizaadeh and Bakhshaei 2009; Bernsdorf and Wang 2009; Ohta et al. 2011; Wang and Bernsdorf 2009). However, Casson model has limited applicability of over a small range of shear rates and the determination of yield stress in the equation is questionable according to Bernsdorf and Wang (2009). To overcome such limitations, present research uses Carreau-Yasuda model for idealizing the rheology of blood.

In the present work, we conduct pore-scale simulation of non-Newtonian blood flow over a range of porous media macroscopic properties. The simulations provide the accurate distribution of velocity, shear rates and viscosity inside the pore-space deciphering the effects of pore microstructure on the flow field. The results, obtained from the above simulations, establish the relationship between equivalent viscosity and porous media shear rate for different pore morphology. Present work further correlates the scaling factor, for the blood flow, with the measurable macroscopic properties of porous media. Such correlations help in the macro-scale simulations of blood flow in porous media. The correlations, proposed in the present work, capture the scaling factor for wide range of porosities including the low porosity values close to percolation threshold. The knowledge obtained from this present computational study paves the way for macroscale hemodynamic simulations in biomedical devices and applications.

## 2 Pore scale geometry

### 2.1 Idealized microstructure geometry reconstruction

Cylindrical solid blocks of circular and square cross sections are arranged in ordered fashion to approximate the pore scale geometry of the porous medium. For ordered arrangement of the blocks, staggered and inline arrangements, as shown in Fig. 1, is considered. For these arrangements, the cases considered include those with solid blocks of (1) only circular cross section, (2) only square cross section and (3) combination of both cross sections. In arrangements of combination of solid blocks with both the cross sections, cases with circular block surrounded by square blocks and cases with square blocks surrounded by circular ones are considered separately, for the better understanding of the effects of the shape of the solid particles on the flow and viscosity distribution.

In all the arrangements, the inclusion ratios of 1:1, 1:2, 1:3, and 1:4 are considered to obtain porous geometries of wide range of porosity and permeability values. Here, inclusion ratio is defined as the ratio of dimension (diameter for circular and side for square) of cylinder at center of the domain to the dimension of cylinder at corner in the case of staggered arrangement. Porosity is varied between 0.4 and 0.9 in all geometries mentioned above except for the geometries for which percolation threshold is higher than the 0.4. Percolation threshold is defined as the value of porosity where all the pores are blocked such that fluid flow no longer takes place.

## 2.2 Stochastic microstructure geometry reconstruction

The results obtained from the simulations, conducted in simplified geometries, are compared with values from simulations in stochastically reconstructed pore microstructure. The complex microstructure of realistic porous materials are reconstructed using the QSGS method devised by Wang et al. (2007). The method is based on the stochastic cluster growth theory developed from pore growth-with-time model by Coveney et al. (1998) in connection with cluster growth theories in Meakin (1998). For the reconstruction of the porous microstructure, the solid matrix of the porous medium is selected as the growing phase and the pores become the non-growing phase(Wang et al. 2013). Initially, the entire volume of the computational domain is assumed to be a pore where any solid phase is absent. The next step is to randomly locate the cores for the growing phase, i.e. the solid phase. For this, all the grid points in the domain are assigned a value in the range (0, 1) randomly, using the random number generator function available in the GNU FORTRAN library. The distribution of the cores is achieved using a core distribution probability, $c_d$, whose value determines the distribution of solid cores throughout the domain which affects the size of the grains of the microstructure. All grid locations having the randomly generated value below $c_d$, will be chosen as a core for the growing solid phase. For a given porosity, larger number of solid phase cores will result in a finer microstructure and vice versa(Wang and Bernsdorf 2009).

Once the cores are assigned, the growth of the porous matrix is governed by the directional growth probability, $D_i$, in the direction i. The value of $D_i$ denotes the probability of a neighboring pore node to get converted into the solid phase. During this step, all the pore nodes neighboring the solid phase cores are assigned random values and the neighbors having $D_i$ values less than the directional growth probability, in their corresponding direction, are converted in to solid phase. The value of $D_i$ in each direction can be adjusted to obtain the required degree of isotropy of the microstructure. This process of conversion of neighboring pore nodes to solid nodes will carry on until the required porosity is achieved.

The QSGS method has been successfully applied in a variety of physical problems(Chen et al. 2015; Chen et al. 2014; Wang et al. 2016). In the present research, QSGS technique is used to generate large numbers of random microstructures of different porosities. In all cases, the porous medium is assumed isotropic by the choice of appropriate directional growth probability values ($D_{1-4}=4D_{5-8}$, where the subscripts are directions defined by Eq. 7). The stochastically reconstructed microstructures often include isolated pores that do not contribute to the bulk fluid flow. For the flow simulation in the reconstructed geometry, all such isolated pores are converted to solid nodes. While such operations slightly modify the porosity of the reconstructed media, the operations allow all the pores to participate in the fluid flow. All subsequent computations, therefore, use porous microstructure free of isolated pores.

## 2.3 Permeability estimation of reconstructed microstructure

Since permeability is a property of the microstructure of the porous media and is independent of the flow and the fluid (Tosco et al. 2013), its value for these geometries, is calculated using Darcy equation by simulating low Reynolds number flow of Newtonian fluid through the geometry. The Darcy law for single phase Newtonian fluid flow in porous media at very low Reynolds number can be written as

$$q = -\frac{K}{\mu}\frac{dp}{dX} \tag{2}$$

Here, $q$ is the specific flow rate, $K$ is permeability, $\mu$ is the viscosity and $\frac{dp}{dX}$ is pressure gradient in the flow direction.

# 3 Governing equations

For Newtonian fluids, shear stress, $\tau_s$, of a fluid is linearly related with shear rate, $\dot{\gamma}$, as $\tau_s = \mu\dot{\gamma}$, where $\mu$ is constant. However, for non-Newtonian fluids, the relation becomes non-linear as the viscosity becomes a function of the shear rate. The stress-strain relationship for non-Newtonian fluids can be approximated as

$$\tau_s = C\dot{\gamma}^n ; \text{ where } \mu = C\dot{\gamma}^{n-1} \tag{3}$$

Here $\mu$ is the apparent viscosity of fluid, $C$ is the consistency constant and $n$ is power law index.

This equation is known as power law model of viscosity or Ostwald-de Waele relationship. It is the basic form adopted to represent viscosity of the non-Newtonian fluid, and a widely used model for non-Newtonian fluid flow simulation due to simplicity of the equation. However, it is valid only for the small range of shear rate, as for low and high shear rates, viscosity will approach to zero or infinity depending on power law index ($n$). Shortcomings of power law model can be overcome by use of Carreau-Yasuda model given by:

$$\frac{\mu - \mu_\infty}{\mu_0 - \mu_\infty} = \left[1 + (m\dot{\gamma})^n\right]^{-a} \tag{4}$$

Where, $\mu_0$ =0.16 Pa.s, $\mu_\infty$=0.0035 Pa.s, $m$ =8.2 s, $n$ =0.64, $a$ =1.23.

$\mu_0$ and $\mu_\infty$ are viscosity of the fluid at zero and infinite shear rate respectively, $m$ is a time scale where $1/m$ represents the critical shear rate at which viscosity turns from Newtonian to shear thinning region. The model can depict viscosity for shear thinning fluid ($n$ <1) correctly even for very low and high shear rate. The values of parameters presented here are for blood and is obtained by curve fitting of experimental data (Wang and Bernsdorf 2009). For $a = 0$ or $m = 0$, Carreau-Yasuda model represent Newtonian fluid with viscosity $\mu_0$. Power law viscosity and Carreau-Yasuda model viscosity with the bulk viscosity limits of 0.16 Pa.s and 0.0035 Pa.s are plotted in Fig. 2. As discussed earlier, for Carreau-Yasuda model, unlike for the power law, at $\dot{\gamma} = 0$ and $\dot{\gamma} = \infty$, the viscosity reaches the respective bulk viscosity values.

Even though models relating local viscosity and local shear rate are available, non-Newtonian fluid in porous media cannot be modeled using the Darcy law in the macroscopic scale due to the non-uniform distribution of viscosity value (Tosco et al. 2013). As discussed in the previous section, in order to use Darcy law to model non-Newtonian fluid flow in porous media, a porous medium viscosity, which represents the equivalent viscosity of fluid when flowing through the porous medium, is to be used. Porous medium shear rate is assumed to be related to the properties of porous medium and the flow rate as given by Eq. (1). Assuming that the porous medium equivalent viscosity ($\mu_{pm}$) and the porous medium shear rate ($\alpha \frac{q}{\sqrt{K\varepsilon}}$) are related according to the Carreau-Yasuda model, it can be written as:

$$\frac{\mu_{pm} - \mu_\infty}{\mu_0 - \mu_\infty} = \left[1 + (m\alpha \frac{q}{\sqrt{K\varepsilon}})^n\right]^{-a} \qquad (5)$$

This relation is made use to conduct a detailed study of scaling factor, by simulating fluid flow through the geometry of known porosity and permeability, for a wide range of pressure gradients and the flow rates. The porous media equivalent viscosity is calculated using the Darcy law, with the pressure drop and flow rate values obtained. Equation 5 is then used for the calculation of scaling factor, $\alpha$, by least square fitting of equivalent viscosity with porous media shear rate.

## 4  Lattice Boltzmann Method

Since the simulation of fluid flow through the narrow passages in the pore scale makes a direct solution of Navier-Stokes equation unstable (Jeong et al. 2006), lattice Boltzmann method (LBM) is used in the present work, for this purpose. LBM (Chen and Doolen 1998; McNamara and Zanetti 1988; Succi 2001; Wolf-Gladrow 2000) refers to a class of computational techniques that solves the Boltzmann equation for an ensemble-averaged distribution of interacting particles over a symmetric, discrete lattice. The particle distribution function is then correlated to the macroscopic variables such as velocity and pressure, thereby, selectively fusing various features of the molecular- to continuum-scale models for practical problem solving. The mesoscopic nature of LBM makes the technique appropriate for simulations of transport in porous media in pore scale.

In LBM, the geometric domain is divided into a regular lattice and groups of molecules are assumed to move through the lattice in their associated directions. This step is called streaming. After streaming, the momentum is redistributed among the molecules through a collision step at each lattice point. The collision step was approximated by Bhatnagar, Gross and Krook (Bhatnagar et al. 1954) with the assumption that all the velocity modes relax towards equilibrium at the same rate. Since the same value of relaxation time is used for all modes of velocity, this approach is called single relaxation time (SRT) approach. Even though, SRT collision is the simplest LBM scheme for flow simulation, it has many limitations rendering it unfit for accurate pore scale simulations. The major drawback is the dependence of wall position on the fluid viscosity which makes the properties of porous media like porosity and permeability a function of the flowing fluid.

Another approach for collision is the Multi-Relaxation Time (MRT) collision which conducts the collision in the moment space such that the parameters used for the various modes can be varied and the hydrodynamic and the kinetic moments can be given separate relaxation times (d'Humieres 1994; Higuera et al. 1989; Lallemand and Luo 2000; Liu et al. 2016). Even though MRT LBM is computationally expensive than SRT, it is shown that the increase in the computational cost is only 10-20% (Lallemand and Luo 2000). Moreover, the viscosity dependence of boundary locations, especially at under relaxed situations (Pan et al. 2006), associated with SRT is overcome by MRT making its use unavoidable in situations where accurate representation of boundary is required (Jithin et al. 2017; Pan et al. 2006).

In the present simulations, lattice Boltzmann equations with MRT collision is used, as shown below:

$$\mathbf{f}(\mathbf{x}+\mathbf{e}\Delta t, t+\Delta t) - \mathbf{f}(\mathbf{x},t) = -\mathbf{M}^{-1}\mathbf{S}\left[\mathbf{m}(\mathbf{x},t) - \mathbf{m}^{eq}(\mathbf{x},t)\right] \quad (6)$$

The LHS is the streaming part in which the particle velocity distribution function, $\mathbf{f}$, at position $\mathbf{x}$ at time $t$ gets streamed to $\mathbf{x}+\mathbf{e}\Delta t$ at time $t+\Delta t$. $\mathbf{e}$ is the discretized lattice velocity vector. In the present work, a D2Q9 model is selected with the velocity being discretized into nine components (Koelman 1991; Qian et al. 1992). The scalar components of the velocity vector $\mathbf{e}$ are defined as:

$$e_i = \begin{cases} (0,0) & i=0 \\ c\left[\cos\left(\frac{i-1}{2}\pi\right), \sin\left(\frac{i-1}{2}\pi\right)\right] & i=1,2,3,4 \\ \sqrt{2}c\left[\cos\left(\frac{i-5}{2}\pi+\frac{\pi}{4}\right), \sin\left(\frac{i-5}{2}\pi+\frac{\pi}{4}\right)\right] & i=5,6,7,8 \end{cases} \quad (7)$$

Here $c=\Delta x/\Delta t$ is the lattice speed and is taken as 1 in lattice units.

The term in the RHS of Eq. (6) is the collision term. The collision is done in the moment space as the transformation matrix $\mathbf{M}$ maps the distribution function $\mathbf{f}$ from the velocity space to the moment space. The transformation matrix $\mathbf{M}$ for D2Q9 model is

$$\mathbf{M} = \begin{bmatrix} 1 & 1 & 1 & 1 & 1 & 1 & 1 & 1 & 1 \\ -4 & -1 & -1 & -1 & -1 & 2 & 2 & 2 & 2 \\ 4 & -2 & -2 & -2 & -2 & 1 & 1 & 1 & 1 \\ 0 & 1 & 0 & -1 & 0 & 1 & -1 & -1 & 1 \\ 0 & -2 & 0 & 2 & 0 & 1 & -1 & -1 & 1 \\ 0 & 0 & 1 & 0 & -1 & 1 & 1 & -1 & -1 \\ 0 & 0 & -2 & 0 & 2 & 1 & 1 & -1 & -1 \\ 0 & 1 & -1 & 1 & -1 & 0 & 0 & 0 & 0 \\ 0 & 0 & 0 & 0 & 0 & 1 & -1 & 1 & -1 \end{bmatrix} \tag{8}$$

The products $\mathbf{m} = \mathbf{M}\mathbf{f}$ and $\mathbf{m}^{eq} = \mathbf{M}\mathbf{f}^{eq}$ are vectors in the moment space, and the nine velocity moments are (Lallemand and Luo 2000; Liu et al. 2014):

$$\mathbf{m} = \left( \rho, e, \zeta, j_x, q_x, j_y, q_y, p_{xx}, p_{xy} \right)^{\mathrm{T}} \tag{9}$$

Here $\rho$ is the fluid density, $e$ is related to the energy and $\zeta$ is related to the square of energy. $j_x$ and $j_y$ are the components of momentum in x and y directions. $q_x$ and $q_y$ are related to the components of energy flux in x and y directions. $p_{xx}$ and $p_{xy}$ correspond to the symmetric and traceless parts of the strain rate tensor. The components of equilibrium moments for flow through porous media is given by (Lallemand and Luo 2000):

$$\mathbf{m}^{eq} = \left( \rho_0, e^{eq}, \zeta^{eq}, j_x, q_x^{eq}, j_y, q_y^{eq}, p_{xx}^{eq}, p_{xy}^{eq} \right)^{\mathrm{T}} \tag{10}$$

$\rho_0$ = mean fluid density = 1 in lattice unit; $e^{eq} = \rho_0 \left[ -2 + 3|\mathbf{u}|^2 \right]$; $\zeta^{eq} = 4\rho_0 \left[ 1 - 3|\mathbf{u}|^2 \right]$;
$j_x = \rho_0 u_x; j_y = \rho_0 u_y; q_x^{eq} = -\rho_0 u_x; q_y^{eq} = -\rho_0 u_y; p_{xx}^{eq} = \rho_0 \left( u_x^2 - u_y^2 \right); p_{xy}^{eq} = \rho_0 u_x u_y$

One of the advantages of the MRT collision is that separate relaxation times can be assigned to each moment. Hence the relaxation matrix, $\mathbf{S}$, is a diagonal matrix with relaxation times for each moment as the diagonal elements. The nine diagonal elements as given in (Chen and Doolen 1998; Pan et al. 2006) are $0, \frac{1}{\tau}, \frac{1}{\tau}, 0, 8\left(\frac{2-1/\tau}{8-1/\tau}\right), 0, 8\left(\frac{2-1/\tau}{8-1/\tau}\right), \frac{1}{\tau}, \frac{1}{\tau}$. This choice of relaxation parameter matrix fixes the wall half way between the lattice points making the wall position independent of viscosity (Lallemand and Luo 2000).

The particle distribution function can be related with the density of fluid and velocity as follows:

$$\rho = \sum_{i=0}^{8} f_i \; ; \; \rho \mathbf{u} = \sum_{i=0}^{8} \mathbf{e}_i f_i \qquad (11)$$

Navier-Stokes equations can be recovered from lattice Boltzmann equation with second order accuracy by Chapman-Enskog expansion. From Chapman-Enskog expansion analysis, it is found that the viscosity of the fluid is related with relaxation time by following relation

$$\nu = \frac{1}{3}\left(\tau - \frac{1}{2}\right)\frac{\Delta x^2}{\Delta t} \qquad (12)$$

Unlike Newtonian fluid, where viscosity does not depend on any parameter except temperature of fluid, shear rate dependence of viscosity in non-Newtonian fluid makes solution tedious. Generally, in conventional CFD techniques, shear rate is calculated using central difference scheme of velocity gradient. However, this scheme may not be efficient for LBM as the velocity needs to be calculated for the velocity gradient calculation. One of the advantages of LBM over other CFD methods for non-Newtonian fluid flow simulation is that the shear rate can easily be calculated with second order accuracy using local values of distribution function (Artoli and Sequeira 2006; Boyd et al. 2007). Moreover, calculation of shear rate from the local distribution function directly, makes the parallelization of the code easier and efficient (Boyd et al. 2007). The value of shear rate at each grid point can be found using the formula

$$\dot{\gamma} = \sqrt{\dot{e}_{\alpha\beta}\dot{e}_{\alpha\beta}} \tag{13}$$

$$\text{Where,} \quad \dot{e}_{\alpha\beta} = -\frac{3}{2\tau}\sum_{i=1}^{n} f_i^{(1)} c_{i\alpha} c_{i\beta} \tag{14}$$

Here, $\dot{\gamma}$ is the shear rate, $\dot{e}_{\alpha\beta}$ is the strain rate tensor and $f_i^{(1)}$ is non-equilibrium distribution function ($f_i^{(1)} = f_i - f_i^{eq}$).

Fluid flow was induced in the domain by applying a pressure gradient in the X direction (Zou and He 1997). No slip at the solid fluid interface was achieved using simple bounce back scheme. Periodic boundary condition is applied in the Y direction. The grid size and time steps are so chosen that relaxation time lies in between 0.71 and 10. As the flow develops, shear rate at each lattice point is calculated from Eq. (13). Viscosity at each point, which is a function of the local shear rate, is found using the Carreau-Yasuda relation given by Eq. (4). The local values of viscosity are implemented in the solution of LBM using the relaxation time which is related to the viscosity according to the Eq. (12). The value of shear rate is calculated using the value of relaxation time at the previous time step.

For each geometry, simulations are performed for a broad range of flow rates ($10^{-10}$ to 0.01 lattice units). Using the Darcy law, equivalent viscosity is calculated from the simulated results for each value of flow rate. The calculated equivalent viscosity is then curve fitted using Eq. (5) in order to calculate the scaling factor. From the simulated flow, the effect of the solid geometry on the distribution of shear rate and viscosity and the flow rate is studied. Attempt is made to quantify the relation between the properties of porous media and the scaling factor.

# 5 Validation

## 5.1 Newtonian Fluid flow in Clear media

For validation of the code, steady flow of a Newtonian fluid past a square cylinder is simulated and compared with the results in published literature (Breuer et al. 2000). For the same, fluid flow is simulated in a computational domain of 500×80 lattice units and a square cylinder, with blockage ratio (ratio of side of square and height of solution domain) of 1/8, is placed at a distance of one third of the length of the domain, from the inlet. Parabolic velocity profile is applied at the inlet. No slip at all solid surfaces is obtained by the use of bounce back scheme (He et al. 1997). The flow outlet boundary condition is applied using the extrapolation scheme for calculating the inward pointing distribution functions. The length of domain downstream of the cylinder is so chosen that the effect of flow outlet could not affect the results. Reynolds number and drag coefficient calculations are carried out using the maximum velocity at inlet as the reference velocity. The coefficient of drag is calculated as:

$$C_D = \frac{F_x}{0.5\rho U_{max}^2} \quad (15)$$

Here, $F_x$ is the x-component of the total force, $\mathbf{F}$, that the fluid exerts on the solid body, which is calculated using the momentum exchange method (Arumuga et al. 2014; Yu et al. 2003) in which the momentum exchange between the solid nodes and the nearest fluid nodes are calculated using the equation:

$$\mathbf{F} = \sum_{\mathbf{x}_b}\sum_{\alpha=0}^{8}\mathbf{e}_\alpha \left[ f_\alpha(\mathbf{x}_b,t) + f_\alpha(\mathbf{x}_b + \mathbf{e}_{\bar{\alpha}}\Delta t, t) \right] \times \left[ 1 - w(\mathbf{x}_b + \mathbf{e}_{\bar{\alpha}}) \right] \Delta x / \Delta t \quad (16)$$

In the above equation, $w(\mathbf{x}_b + \mathbf{e}_{\bar{\alpha}})$ is an indicator function with value 0 at the fluid nodes and 1 otherwise. Also, $\bar{\alpha}$ is the direction opposite to $\alpha$. The force is calculated by summing the total momentum exchange between each surface node and the adjacent fluid node over the entire surface area of the solid body.

Cases with Reynolds numbers of 10, 15, and 30 are simulated and the streamlines for each of these cases are shown in Fig. 3. Drag coefficient for each case is calculated and compared with values given by Breuer et al. (2000), obtained using FVM and LBM simulations, in Table 1. The results are in good agreement with Breuer et al. (2000), showing the correctness of the simulations.

## 5.2 Newtonian Fluid flow through porous media

Validation of fluid flow through porous media is conducted by simulating flow of Newtonian fluid through a porous geometry assumed by array of circular cylinders. The flow is induced by imposing a pressure gradient such that the Reynolds number of flow is very low and the Darcy law is valid. From the known values of imposed pressure gradient and porosity and the measured value of volumetric flow rate, permeability of the geometry is calculated using the Darcy law.

Values of permeability obtained from simulations are compared with correlations by Lee and Yang (1997) and Gebart (1992) as shown in Fig. 4a. Both Lee & Yang correlation and Gebart correlations were developed with the assumption of porous media as an array of regularly ordered parallel fibers, similar to the assumption made in the present work. The figure shows plots of permeability, non-dimensionalized with square of diameter of the cylinder, and porosity of the array of cylinders. The figures show that the permeability values calculated from the simulations are in excellent agreement with the correlations, showing the capability of the present simulation technique in handling fluid flow in porous media of complex geometry.

**5.3 Non-Newtonian fluid flow**

To validate the simulation of non-Newtonian fluid flow, power law fluid is assumed to flow through a complex pore geometry used by Sullivan et al. (2006). Simulations of different fluids were conducted by varying the power law index, $n$. From the simulations, the flow rates are measured for different pressure gradients applied. According to power law, the flow rate and pressure gradient are related according to

$$q = -K'\left(\frac{\Delta p}{l}\right)^{1/n} \tag{17}$$

Here, $l$ is the distance of the computational domain along the flow direction and $K'$ is a constant which depends on the porous medium and the flowing fluid (Sullivan et al. 2006). According to Eq. (17), the plots of $\log(q)$ vs. $\log(\Delta p/l)$ should be straight lines with slope $1/n$. Figure 4b shows plots of $\log(q)$ vs. $\log(\Delta p/l)$ for different fluids. The plots show straight lines with different slopes for the different fluids. Slope of these curves were calculated using least square fitting and the values are listed in Table 2. The table shows the values of slopes calculated and the values of power law index and the percentage error in each case. The error for all cases falls less than 2%. Hence it was concluded that our code could successfully handle non-Newtonian fluid flow in complex geometry porous media accurately.

# 6 Results and Discussion

## 6.1 Permeability Estimation

Permeablity of porous media is one of the most important parameters in the study of flow in porous media. It indicates the ability of the porous medium to allow fluid to flow through it. In the case of non-Newtonian fluid flow, permeability has important role in the calculation of porous medium equivalent viscosity and porous medium shear rate.

For all the geometries generated, permeability is calculated using Darcy law by simulating low Reynolds number flow of Newtonian fluid. Plots of permeability, non-dimensionalized with square of the hydrodynamic radius (4A/P), with porosity for different geometries are shown in Fig. 5. Figure 5a and 5b show the permeability variation for arrangement of circular blocks and square blocks, respectively. Figure 5c is the plot for circular block surrounded by square blocks (denoted by 'square-circular arrangement' from here onwards) and Fig. 5d is for square cylinder surrounded by circular blocks (denoted by 'circular-square arrangement' from here onwards). In all the cases, permeability is zero for porosity below the percolation threshold. At porosity values just above the percolation threshold, the rate of increase of permeability with porosity is high. As the porosity is further increased, the increase in permeability becomes more gradual. Among the different arrangements, for inline arrangement of square blocks, the rate of increase of permeability with porosity is smaller for lower porosities (Fig. 5b). This is because of the smaller value of percolation threshold porosity of this geometry as can be seen from Fig. 6a. For this geometry, the percolation threshold is reached when porosity is zero.

From Fig. 5b, 5c and 5d it can be seen that the value of permeability for lower porosities vary significantly, especially for geometries involving square cylinders. This can be explained from Fig. 6 which shows the comparison of flow field of inline and staggered arrangements of circular and square blocks. From the figure, it can be seen that the constriction of flow passage is the factor limiting the flow rate and the narrowness of flow passage depends on the shape of the solid blocks. If we compare the staggered arrangement of circular and square cylinders, it is seen that the constriction of flow passages is more for square blocks for the same porosity. This means that the percolation threshold will be higher for square block arrangement. The effect of constriction is profound especially when the porosity is near the percolation threshold. It can be suggested that this constriction is being reflected in the values of permeability. The change in area of flow also leads to lager velocity gradients resulting in larger strain rates which lead to a reduction in viscosity in those regions for shear thinning fluids. Hence it is inferred that the change in viscosity in the case of non-Newtonian flow in porous media is related to the value of permeability of porous media. This assumption is made use in the coming sections of the paper while generating a relationship of the scaling factor with the properties of the porous media.

### 6.2 Non-Newtonian Fluid Flow Simulation

Flow of non-Newtonian fluid is simulated in the porous geometries and the contours of normalized velocities and streamlines of flow are shown in Fig. 7. The figure includes results for three different values of porosity (0.3, 0.6, and 0.9) of porous media of circular cylinders arranged in inline and staggered fashion. The plots show that in geometries with larger porosity values (Fig. 7c and 7f) the variation of velocity in the bulk of the fluid is small when compared to that in low porosity geometries. This is due to the spatial separation of large parts of the flow field from the solid surface. This, in turn, results in the smaller variation of strain rate and a more uniform viscosity distribution as seen from Fig. 8c and 8f. However, in the case of inline cylinders, as the particle sizes are increased to lower the porosity, the regions of secondary flow are observed in spaces between the circular blocks (Fig. 7a). These regions have very small velocities and the velocity gradient values are also seen to be minimal. Due to the low values of shear rate, the viscosity in this region is the highest (Fig. 8a). On the other hand, the flow velocity gradient is observed to be the maximum in regions between cylinders where the flow cross sectional area is constricted. In these regions, the strain rate is maximum resulting in the decrease in viscosity.

Very similar trends are observed when the inclusions are square cylinders. The plots of viscosity distribution for inline and staggered arrangement of square cylinders are shown in Fig. 9. It can be seen that for high porosity geometry the distribution of viscosity is almost uniform (Fig. 9c and 9f). In the case of low porosity of inline square cylinders (Fig. 9a), the regions of secondary flow are larger with almost zero velocity. The low values of shear rates arising there, results in large values of viscosity in those regions. For staggered arrangement of square cylinders, the flow passage constriction near the corners of the cylinders is more abrupt and localized making the velocity gradients at that region more profound. The high strain rate values result in a sharp decrease in the viscosity near these corners as can be seen from Fig. 9d. Simulations are conducted for square-circular and circular-square arrangements and results of viscosity distribution are shown in Fig. 10. In these cases also, it is seen that for high porosities the distribution of viscosity is more or less uniform. For smaller porosities, variation of viscosity is seen to be profound near the sharp corners arising due to the presence of square cylinders. In these cases also the viscosities are seen to decrease at the flow passage constrictions between cylinders.

### 6.3 Newtonian vs. Non-Newtonian

To closely observe the effect of viscosity variation on the flow, X-direction velocity profiles along the Y-direction is plotted in the pores between cylinders, at X=0, for different cases for both Newtonian and non-Newtonian fluids. In all the figures, the velocity is normalized with the velocity midway between the solid surfaces in the domain at that section. Figure 11a shows the velocity profile between two circular cylinders in the inline arrangement for Newtonian and non-Newtonian fluids. The dashed lines show velocity profiles of Newtonian fluid. It can be seen that the velocity gradient in the direction normal to the solid surface is smaller for Newtonian velocity profile when compared to that of non-Newtonian fluid. This is because of the reduction in viscosity of non-Newtonian fluid near the solid surface due to the strain rate. However, the viscosity in the bulk of the fluid away from the solid is still high when compared to that near the wall. This distribution of viscosity results in a velocity profile, as seen in the plot, with more uniform velocity distribution away from the wall.

Similar profiles are plotted for staggered arrangements of cylinders of circular, square and combination of both cross sections. It can be seen that there is a reduction of velocity at the midpoint when square cylinder is at the center of the domain. This is due to the larger obstruction to flow delivered by the square cross section. Plots of velocity profiles for circular and square inclusions with different size ratios are also plotted for comparison (Fig. 11c and 11d). All these cases are seen to show similar trends.

### 6.4 Equivalent Viscosity and Scaling Factor

Equivalent porous medium viscosity is calculated using modified Darcy law (Eq. (2)) for different geometries and flow rates. The calculated equivalent viscosity is plotted with flow rate and porous medium shear rate and is shown in Fig. 12. In the figure, the equivalent porous medium viscosity is plotted for square and circular cylinders for different arrangements and inclusion ratio. As we can see from the figure, porous medium viscosity varies with flow rate and the nature of plots are similar to the curve predicted by Carreau-Yasuda model. The figure suggests that for a flow rate, the value of porous media equivalent viscosity changes with the pore geometry. It proves that relation of equivalent viscosity with flow rate is not universal and rather it will change with geometry. On the other hand, the equivalent viscosity is plotted with porous medium shear rate (Eq. (1)) in Fig. 12d. As we can see from this plot, for all the cases, equivalent viscosity value lies on the same line. This shows that the relation between the porous media viscosity and porous media shear rates are not dependent on the micro-scale properties.

However, the porous media equivalent shear rate plotted in Fig. 12d is different from the definition of shear rate obtained by dimensional analysis ($q/\sqrt{K\varepsilon}$) which does not predict the correct value of equivalent viscosity of flow. In order to match the viscosity value predicted with the correct value of equivalent porous media viscosity, the scaling factor is introduced in porous medium shear rate (Eq. (1)). This can be seen from the Fig. 13a, where viscosity, obtained from present simulations and Carreau-Yasuda model with and without use of scaling factor, is plotted with the porous medium shear rate. It can be seen from the figure that without scaling factor, Eq. (5) over predicts viscosity and a scaling factor has to be used to predict viscosity correctly. It is also important to note that Eq. (5) predicts viscosity correctly even without using scaling factor at low and high shear rate proving that scaling is not required in Newtonian regime.

The importance of scaling factor in macroscopic study of flow can be appreciated from Fig. 13b. In this plot, the pressure difference between inlet and outlet is plotted with the flow rate for Carreau-Yasuda model fluid for porosity value of 0.4 and 0.9. Pressure drop is calculated at a given flow rate for the equivalent viscosity predicted by LBM simulations and the value of viscosity predicted without scaling factor. It is clear from the figure that for low and high values of flow rate, pressure gradient is matching with actual value, but not in the intermediate range of flow rate, where pressure difference is overestimated. The reason for this is that, for low and high values of flow rate, the shear rates will also have respective low and high values. Under these conditions, the spatial viscosity variations are minimal and the flow resembles to that of a Newtonian fluid. This results in the correct prediction of pressure difference at high and low flow rates. For the intermediate range, however, the non-Newtonian behavior is prominent causing non-uniform distribution of viscosity, causing the inaccurate prediction of pressure difference without the use of scaling factor. This deviation of pressure difference from the actual value can give rise to error in macroscopic calculations. From this, we can conclude that, even though, for very low or high flow rates, the scaling factor does not play any role, intermediate ranges of flow rate require scaling factor to be used for accurate flow representation.

## 6.5 Scaling Factor and Pore Geometry

In this section we attempt to find out how the value of scaling factor varies with different geometries of pore structure. The major parameters of microstructure of the porous media affecting the scaling factor are studied here. For this purpose, plots of scaling factor and the macroscopic properties of porous media have been made and discussed. However, it should not be perceived that the plots are made by varying one property while keeping other parameters constant, as it is not possible to vary a single property of the porous medium without affecting the others. All the plots are shown for qualitative understanding of dependence of scaling factor on the properties of porous media.

In Fig. 14, the scaling factor is plotted with porosity for different geometries. The scaling factor for staggered arrangement of square cylinders, circular cylinders, and their combinations for different inclusion size ratios have been compared. It is observed that for all the geometries, the scaling factor value is minimum and more or less constant for high porosities. This is because of the more uniform distribution of viscosity in the entire domain for high porosity, as discussed in the previous section. However, for low porosity geometries, the scaling factor values are large. This is due to the narrowness of the flow passages which results in the larger variation in the cross sectional area for flow, giving rise to larger variations in strain rate. This will result in a very non-uniform distribution of viscosity in the domain. It can be presumed from this discussion that the scaling factor takes large values as the variations in the local viscosity values in the domain become significant. Since the variations in local viscosity increases as the porosity is decreased, the scaling factor value also increases and reaches its maximum when the percolation threshold is reached.

It is also seen from the figures that different geometries obtained by varying the solid block shapes and inclusion size ratios have different porosity values of maximum scaling factor. This can be analyzed based on the percolation threshold of each geometry given in Table 3. The table shows that the square inline geometry is the one with minimum percolation threshold which is zero. The curve of scaling factor for this geometry (Fig. 14b) shows that the value is almost constant for all porosity values simulated. Excluding the inline square geometry, circular cylinder arrangements have the lowest percolation threshold when compared to other geometries and the staggered square has the highest. If these two geometries are compared, for same inclusion ratio, the scaling factor for square staggered is much larger for the same porosity. The square-circular and the circular-square geometries have percolation threshold values between the square staggered and circular staggered. If we compare the scaling factor values of inclusion ratio 1:1 for these geometries, it is seen that the values of square-circular and circular-square geometries lie in between the values of circular staggered and square staggered.

Moreover, it is seen that as the inclusion size ratio is increased for a particular arrangement, the percolation threshold value is seen to decrease. From Fig. 14, it is seen that the scaling factor value also decreases with increasing inclusion size ratio. From the discussion it is suggested that the scaling factor has a direct correlation with the percolation threshold of the geometry and the dependence is strong except for a small range of very high porosity where the changes in flow passage narrowness become insignificant.

## 6.6 Scaling Factor and Macroscopic properties of Porous Media

In spite of the importance of scaling factor, there is no correlation available to calculate it. In order to find a relation for the scaling factor, its dependence on macroscopic parameters needs to be known. In literature, it is mentioned that the scaling factor depends on properties of porous media and the rheology of the fluid (Sorbie et al. 1989; Tosco et al. 2013). But, the properties on which it depends have not been specified clearly. Some authors mentioned scaling factor is affected by properties of porous media like the tortuosity(Lopez et al. 2003). But, a detailed study of the relationship of scaling factor with these parameters is missing in the literature, especially for fluids of complex rheology. Presently, it is considered that the scaling factor is dependent on the porosity, tortuosity, permeability and the percolation threshold. Permeability and the percolation threshold values are selected according to the premise from the discussions in the previous sections. From the trends observed in the plots discussed, it is considered that as the porosity tends to percolation threshold, the value of scaling factor tends to very high values. Hence the effect of percolation threshold and porosity is included in the correlation as the difference between the two.

It has been discussed by Pearson and Tardy (Pearson and Tardy 2002) that the fluid tends to flow through the least tortuous path and that the tortuosity has an important effect on the value of the apparent viscosity. Also, from physical reasoning, it is inferred in the present work that the more tortuous the flow paths are, the more will be the variations in the values of the local viscosity. As discussed in previous section, the variation in viscosity values is seen to have strong relation with the value of scaling factor. Based on all these factors, tortuosity is also considered in the development of the correlation.

The relation between scaling factor and these macroscopic properties of porous media are studied and attempt is made to quantify the relation for blood flow in simple geometries.

In this regard, assumption of a correlation is made, between the scaling factor and the macroscopic properties, of the form:

$$\alpha = c\left(\frac{K}{L^2}\right)^x T^y \left(\varepsilon - \varepsilon_p\right)^z \qquad (18)$$

Here, $K$ is the permeability, $\varepsilon$ is the porosity, $T$ is the tortuosity and $\varepsilon_p$ is the porosity at percolation threshold. $L$ is the hydraulic radius of the solid block used to non-dimensionalize the permeability. Carreau-Yasuda model is used for the shear thinning behavior of the fluid. Equation 18 is least square fitted to calculate the value of $c$, $x$, $y$ and $z$. The values of these parameters are shown in Table 4.

From the values of these constants, the nature of relationship between scaling factor and the porous media properties can be understood. The correlation suggests that the scaling factor increases with permeability but only weakly when compared to other parameters, since the power to which permeability is raised (x) is of the order of 0.1. The values of y and z are negative showing that the scaling factor is inversely proportional to the tortuosity and the difference between the porosity and the percolation threshold porosity. The relationship is close to linear as $y$ and $z$ are of the order 1. The trend followed by the scaling factor with porosity can be explained using this correlation. For very low porosity, the value of scaling factor becomes very high and goes to infinity for percolation threshold porosity, i.e. $\varepsilon = \varepsilon_p$. As the porosity increases from percolation threshold, the $\varepsilon - \varepsilon_p$ term increases, reducing the value of scaling factor. At the same time, increased porosity enhances permeability and that may increase the scaling factor. The above two competing effects leads to a minima of scaling factor at high porosity (~0.8) followed by slight increase of the factor. The phenomenon is efficiently captured in Fig. 14. The increase in the scaling-factor near the maximum porosity indicates the influence of permeability on the scaling factor.

## 6.7 Scaling Factor for Realistic Porous Media

The correlation, proposed in the previous Section, provides basic understanding on the dependence of scaling factor on the porous media properties. Such correlation, derived over idealized porous geometry requires rigorous testing flow through real porous media. Present study, therefore, extends the pore-scale simulation of blood flow over stochastically reconstructed porous media. The reconstruction technique generates random porous media of varied porosities using the QSGS method. Permeability of the reconstructed media are calculated using the technique explained in Section 6.1. The contours of viscosity and velocity for non-Newtonian fluid flow in complex reconstructed microstructures with porosities 0.8 and 0.7 are shown in Fig 15. The contours, shown in Fig 15, indicate significant variation in viscosity over the domain. The local viscosity values attain their minimum at the regions of high flow rates while maximum viscosities are realized at the stagnation regions. Such trend in viscosity variation remains independent of porosity, as indicated in Fig 16a. The trends in local viscosity variation also match with the same for flow through idealized geometries, as shown in Fig 8-10. Also, for the same flow rate, the equivalent viscosity value is different for different pore geometries. However, when plotted with the equivalent porous medium shear rate, the equivalent viscosity values became independent of the porosity, as seen in Fig 16b. The Carreau-Yasuda model, without the scaling factor, however, still overpredicts the equivalent viscosity for the intermediate shear rate where the fluid behaves as non-Newtonian. The comparison of simulation results with the prediction of Carreau-Yasuda model with and without scaling factor is shown in Fig 16b. Such trend mimics the same in the idealized geometries, as shown in Fig 13a. The Carreau-Yasuda model, therefore, requires a scaling factor for the accurate prediction of viscosity values both for stochastic as well as idealized geometries.

The applicability of the correlations, proposed in the present work, is investigated by comparing the values of scaling factor obtained from the correlation to the same calculated from the non-Newtonian flow simulations. To develop the correlation, the properties of the stochastic microstructures, namely the porosity, permeability and tortuosity, are evaluated from the pore-scale simulations, as discussed in Section 6.6. The values of permeability for the different pore geometries are plotted against porosity in Fig 17a. Following similar trend as in the idealized geometries, computed permeabilities, for the high-porosity stochastic geometry, match quite well with the predictions by Lee and Yang (1997) and Gebart (1992). For the low-porosity random microstructures, however, the computed permeabilities show relatively larger deviations. For the stochastic reconstruction of the porous media, several microstructures are generated for the same porosity. The resulting plots, shown in Fig 17a and 17b, indicates the mean from several simulations. The error bars, shown in Fig 17a and 17b indicate 98% confidence interval.

To calculate the scaling factor $\alpha$, the values of permeability, porosity and tortuosity of the realistic microstructures are used as the inputs to the Eq (18). The values of $\alpha$, thus obtained, are compared with the simulation results as shown in Fig 17b. The predictions of the correlations show excellent agreement with the scaling factor values obtained from simulations. The variation in the value of scaling factor for the microstructures of high porosities, is found to be negligible. Moreover, for higher porosities, the computed values $\alpha$ are captured very efficiently by the proposed correlation. As the porosities decrease, especially when the porosities are close to the percolation threshold, the estimated scaling factors, however, show high uncertainties. Near the percolation threshold, small variation in the microstructure induces to the large changes in the flow field leading to the uncertainties in the scaling factor prediction. Overall, Fig 17b confirms the accuracy of the correlations obtained for a variety of porous microstructures. Compared to other microstructures, however, the square blocks geometry, show slight under-prediction in the scaling factor. Such findings indicate that the grain shape of the porous media also plays a role in the accurate prediction of scaling factor and underscores the requirement of further studies to quantify the influence of the grain shape on the scaling factor.

From the values of scaling factor obtained from random pore structures, a correlation of the form of Eq. (18) has been developed and the coefficients and powers of the different parameters are included in Table 4. This correlation suggests that the scaling factor is inversely related to tortuosity and the difference between porosity and percolation threshold. Further, the permeability dependence of scaling factor, for the random microstructure, shows similar trend to that obtained from the idealized microstructure. The correlation, for the random microstructure, may further be improved by considering multiple inputs such as grain geometry, pore connectivity and pore size distribution during stochastic reconstruction.

Furthermore, the constants, appearing in the Eq. 18, vary for each geometry suggesting that the proposed correlation is not universal. However, the correlation identifies the properties of porous medium that influence the scaling factor and provides a qualitative understanding of the relationship between the scaling factor and the porous media properties. Further studies are, therefore, necessary to arrive at a universal correlation of scaling factor.

# 7  Conclusions

Present work involves pore-scale simulations of Newtonian and shear thinning non-Newtonian fluid flow using multi-relaxation time lattice Boltzmann method. The goal is to use the computed flow variables to calculate the equivalent viscosity of the non-Newtonian fluid. Finally, the present study attempts to find meaningful correlation of the scaling-factor with the properties of the porous media. Major conclusions from the present studies are as follows:

1. The variation of permeability value with porosity is strongly related to the percolation threshold of the porous media. Small change in porosity, for values near percolation threshold, can cause significant variations in permeability values due to the significant flow cross sectional area constriction. Since such changes in cross sectional area affect the local shear rates and viscosity, it is suggested that the permeability value is related to the extent of non-Newtonian behavior of fluid.
2. For non-Newtonian fluid flow in porous media at high porosities, the variations in viscosity in the pore space is small due to the spatial separation of large parts of the flow field from the solid surface. For smaller porosities, the viscosity distribution becomes highly non-uniform due to the high variations in the local shear rate. The regions of secondary flow in the domain, where the shear rate is minimal, is observed to have the highest viscosity as expected, due to the shear thinning nature of the fluid. Regions of flow passage constrictions between solid grains have larger shear rate causing decrease in the viscosity at those regions.

3. Velocity gradient at solid surface in the normal direction is smaller for Newtonian fluids when compared to that of shear thinning non-Newtonian fluids. This is because of the reduction in viscosity of the shear thinning fluid near the solid surface due to the higher strain rate near solid. This distribution of viscosity results in a more uniform velocity distribution away from the wall for these non-Newtonian fluids.

4. Macroscopic calculations of non-Newtonian fluid flow in porous media require the definition of a porous media equivalent viscosity and a porous media shear rate, related according to a modified Darcy law. However, the calculation of this porous media shear rate requires a scaling factor whose value depends on the properties of porous media and the flowing fluid. Without scaling factor, the porous media shear rate over predicts the equivalent viscosity except for very low and very high shear rates. In macroscopic calculations, this will result in under prediction of flow rates for a given pressure gradient in the intermediate range of flow rate, even though, for very high and very low flow rates it is not affected significantly.

5. The value of scaling factor is affected by the microstructure of the porous media. It is presumed that the scaling factor takes large values as the variations in the local viscosity values in the pores become significant. Since the variations in local viscosity increases as the porosity is decreased, the scaling factor value also increases and reaches its maximum when the percolation threshold is reached. Present study suggests that the scaling factor has a direct correlation with the percolation threshold of the geometry and the dependence is strong except for a small range of very high porosity where the changes in flow passage narrowness become insignificant.

6. Present study proposes a correlation to relate scaling factor with the macroscopic properties of the porous media. The proposed correlation suggests that the scaling factor varies with permeability, but only weakly when compared to other parameters. Further, the scaling factor is inversely related to the tortuosity and the difference between the porosity and the percolation threshold. The scaling factor attains high values at low porosities when the porosity is close to percolation threshold porosity. The scaling factor again realizes high values at very high porosities, a phenomenon attributed to the high values of permeability.

7. The proposed correlations have been successfully extended for flow through stochastically reconstructed porous microstructures. The geometry dependence of the correlation parameters indicates that the scaling factor may depend on additional microstructural characteristics, such as grain geometry, pore connectivity and pore size distribution.

# Nomenclature

| | |
|---|---|
| $c$ | Lattice speed |
| $C$ | Consistency constant |
| $\mathbf{e}, e_i$ | Discretized lattice velocity vector and its components |
| $\dot{e}_{\alpha\beta}$ | Strain rate tensor |
| $\mathbf{f}$ | Vectors of lattice Boltzmann distribution function |
| $K$ | Permeability (m$^2$) |
| $m$ | Critical shear rate (1/s) |
| $n$ | Power law index |
| $\mathbf{m}$ | Vector of distribution function in moment space |
| $\mathbf{M}$ | Transformation matrix |
| $P$ | Pressure (Pa) |
| $q$ | Volumetric flow rate (m$^3$/s) |
| $\mathbf{S}$ | Relaxation matrix |
| $t$ | Time (s) |
| $T$ | Tortuosity |
| $u, v$ | Velocity components (m/s) |
| $\mathbf{u}$ | Macroscopic velocity vector |
| $\mathbf{v}$ | Temporal velocity vector |
| $\mathbf{x}$ | Position vector |
| X, Y | Cartesian coordinate directions (m) |
| $\alpha$ | Scaling factor |
| $\dot{\gamma}$ | Shear rate (1/s) |
| $\dot{\gamma}_m$ | Porous medium shear rate (1/s) |
| $\varepsilon$ | Porosity |
| $\varepsilon_p$ | Porosity at percolation threshold |
| $\tau_s$ | Shear stress (Pa) |
| $\mu$ | Bulk viscosity of fluid (Pa.s) |
| $\mu_\infty$ | Bulk viscosity at infinite shear rate (Pa.s) |
| $\mu_0$ | Bulk viscosity at zero shear rate (Pa.s) |
| $\mu_{pm}$ | Porous medium equivalent viscosity (Pa.s) |
| $\tau$ | Non-dimensional relaxation time |
| $\nu$ | Kinematic viscosity (m²/s) |
| $\rho$ | Density (kg/m³) |

# List of Figures

| | |
|---|---|
| | medium viscosity with porous medium shear rate for the circular geometry |
| **Figure 13** | Porous medium equivalent viscosity v/s porous medium shear rate is plotted from LBM simulation and Carreau-Yasuda model with and without shift factor |
| **Figure 14** | Scaling factor is plotted with porosity for the different geometries for various inclusion size ratios. The insets in each figure show the zoomed view of the curves in the high porosity range. The increase in the value of scaling factor for high porosity can be observed from these insets. |
| **Figure 15** | (a) Contours of dynamic viscosity (SI units) over the domain for porosity=0.8 (b) Contours of X-direction velocity (lattice units) over the domain for porosity=0.8 (c) Contours of dynamic viscosity over the domain for porosity=0.7 (d) Contours of X-direction velocity over the domain for porosity=0.7 |
| **Figure 16** | (a) Variation of equivalent porous medium viscosity with flow rates in geometries of various porosities (b) Comparison of equivalent porous medium viscosity values and their variation with porous medium shear rate between simulation and Carreau-Yasuda (CY) equation |
| **Figure 17** | (a) Permeability with porosity plot for reconstructed geometries compared with simplified geometries. (b) Scaling factor values predicted by the proposed correlations compared with values obtained from LBM simulations |

**Table 1: Drag coefficients for flow past square cylinder compared with results of Breuer et al. (2000)**

| Reynolds number | $C_D$ (Breuer - FVM) | $C_D$ (Breuer-LBM) | $C_D$ (Present study) |
|---|---|---|---|
| 10 | 3.6275 | 3.3644 | 3.3862 |
| 15 | 2.8481 | 2.6997 | 2.7265 |
| 30 | 2.0187 | 1.9721 | 2.1134 |

**Table 2: Comparison of power law index used and calculated from result for validation case**

| Power law index (n) | Slope(1/n) | % error in slopes |
|---|---|---|
| 1 | 1 | 0 |
| 0.7 | 1.408 | -1.4 |
| 0.5 | 1.956 | -1.75 |

**Table 3: Percolation thresholds of different geometries considered**

| Geometry | Arrangement | Inclusion ratio | Percolation threshold |
|---|---|---|---|
| Square | Inline | - | 0.0000 |
| | Staggered | 1/1 | 0.5000 |
| | Staggered | 1/2 | 0.4444 |
| | Staggered | 1/3 | 0.3750 |
| | Staggered | 1/4 | 0.3200 |
| Circular | Inline | | 0.2146 |
| | Staggered | 1/1 | 0.2148 |
| | Staggered | 1/2 | 0.1276 |
| | Staggered | 1/3 | 0.1273 |
| | Staggered | 1/4 | 0.1655 |
| Circular-square | Staggered | 1/1 | 0.3874 |
| | Staggered | 1/2 | 0.3470 |
| | Staggered | 1/3 | 0.2880 |
| | Staggered | 1/4 | 0.2424 |
| Square-circular | Staggered | 1/1 | 0.3874 |
| | Staggered | 1/2 | 0.2894 |
| | Staggered | 1/3 | 0.1718 |
| | Staggered | 1/4 | 0.1521 |

Table 4: Values of correlations for different geometries

|   | Circular | Circular-Square | Square-Circular | Square | Realistic Microstructures |
|---|---|---|---|---|---|
| $c$ | 1.89832 | 1.58336 | 1.69825 | 1.36088 | 1.9417 |
| $x$ | 0.06808 | 0.13885 | 0.090321 | 0.1107 | 0.10199 |
| $y$ | -0.24772 | -1.86295 | -1.02272 | -3.24557 | -2.9758 |
| $z$ | -0.46296 | -0.99527 | -0.703097 | -1.01082 | -0.96489 |

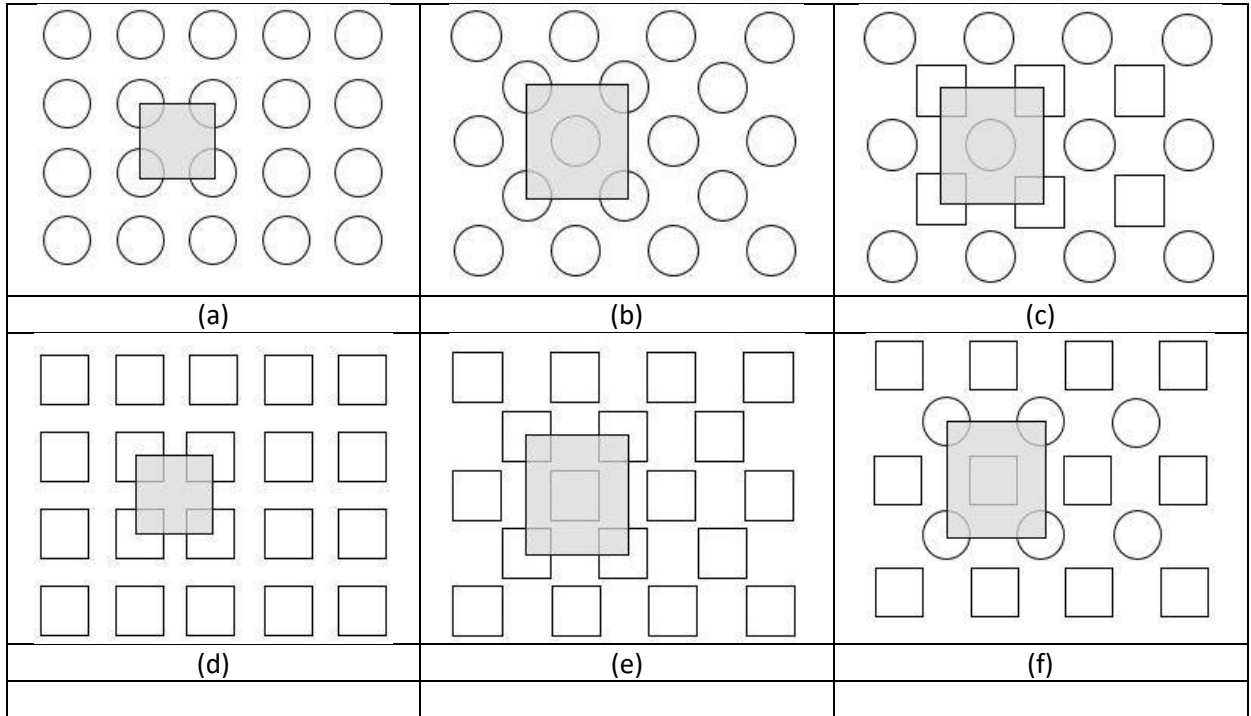

**Figure 1: Pore scale geometries used for the simulation. (a) and (d) are inline arrangement of circular and square cylinders respectively. (b), (c), (e), and (f) represent staggered arrangements. (c) and (f) are geometries where blocks of both circular and square cross sections are present. The dark rectangle shows the unit cell of computations. Cases with computational domain as shown in (c) and (f) are called square-circular arrangement and circular-square arrangement, respectively. (g) and (h) are complex geometries generated using QSGS method before removing isolated pores. The two figures show boundary conditions of the two test cases run to locate the isolated pores. (i) shows the isolated pores and the geometry obtained after their removal.**

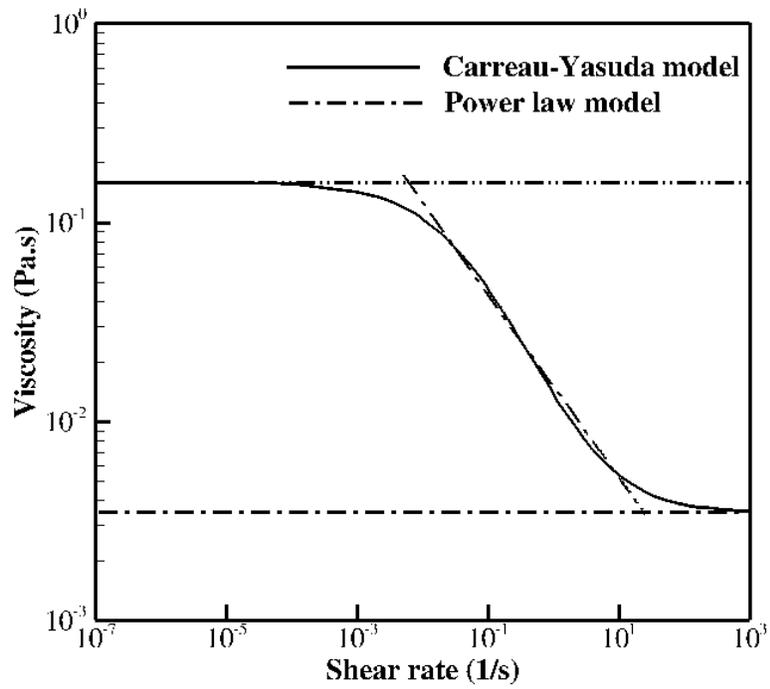

**Figure 2: Comparison of variation of viscosity with shear rate for power law and Carreau-Yasuda model with bulk viscosity limits of 0.16 Pa.s and 0.0035 Pa.s**

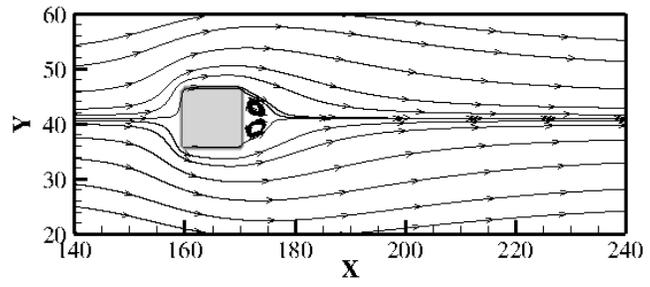

(a)

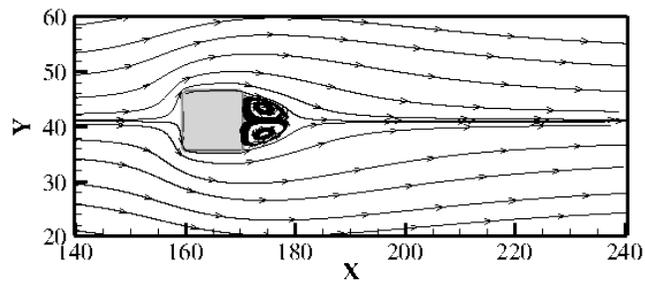

(b)

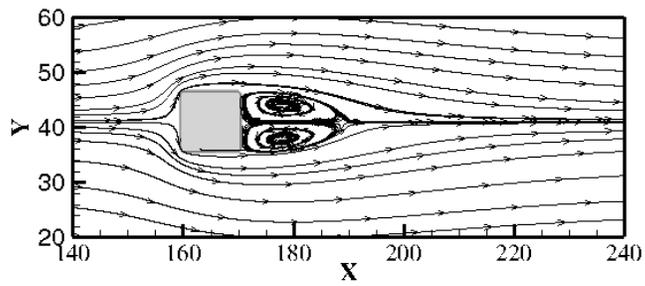

(c)

**Figure 3 : Streamline plots for flow past a square cylinder at (a)Re=10 (b)Re=15 (c)Re=30**

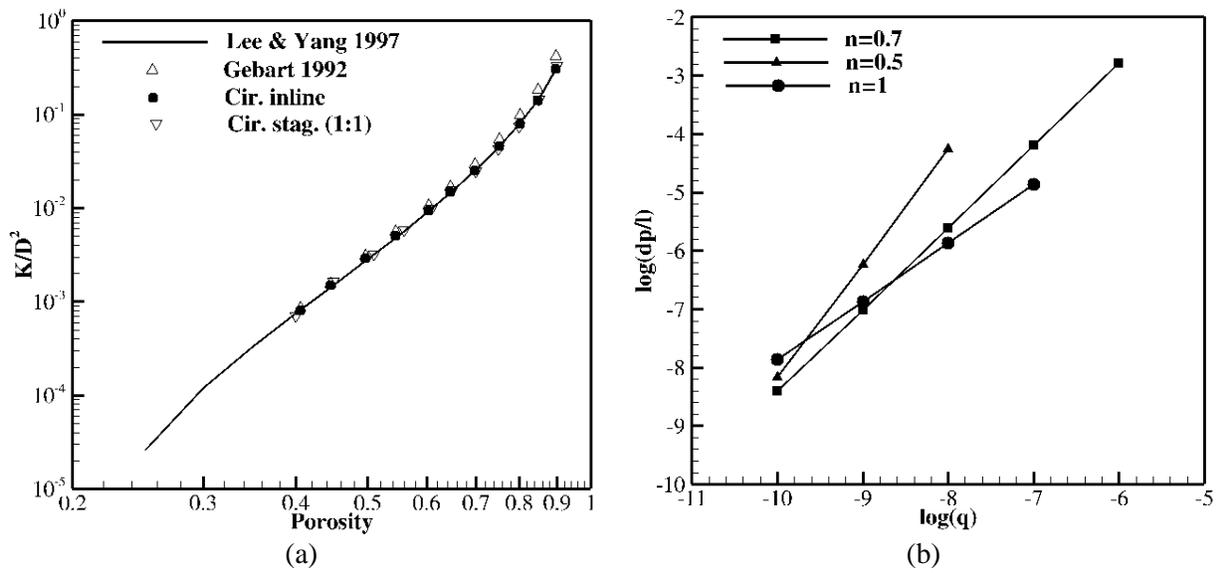

**Figure 4: (a) Dimensionless permeability is plotted with porosity for inline and staggered arrangement of circular cylinder. (b) Force flux law is plotted for power law fluid in a complex geometry.**

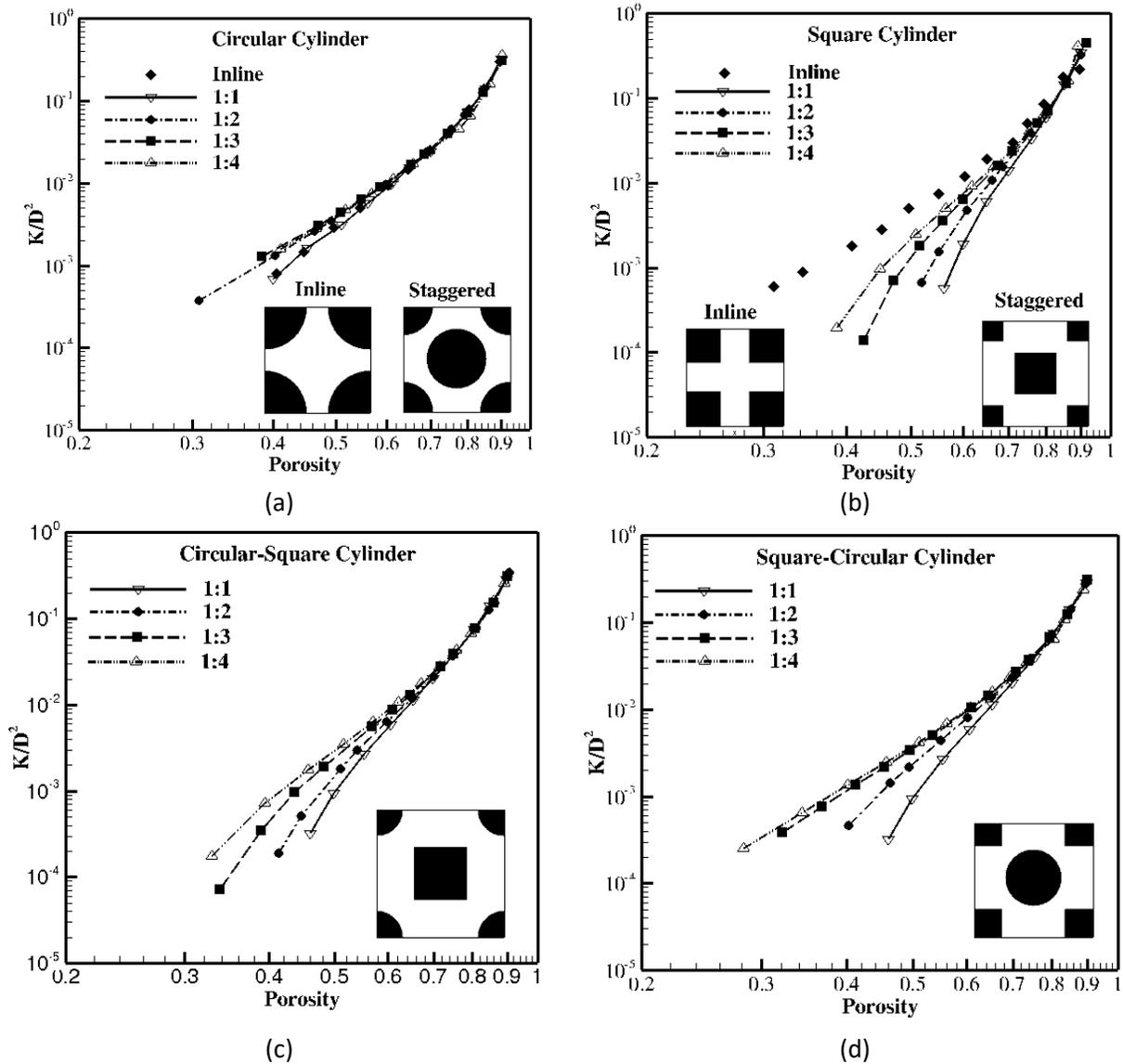

Figure 5: Permeability v/s porosity curve from values obtained from Newtonian fluid simulation for different combinations of geometry. Figures in insets show the geometry of consideration.

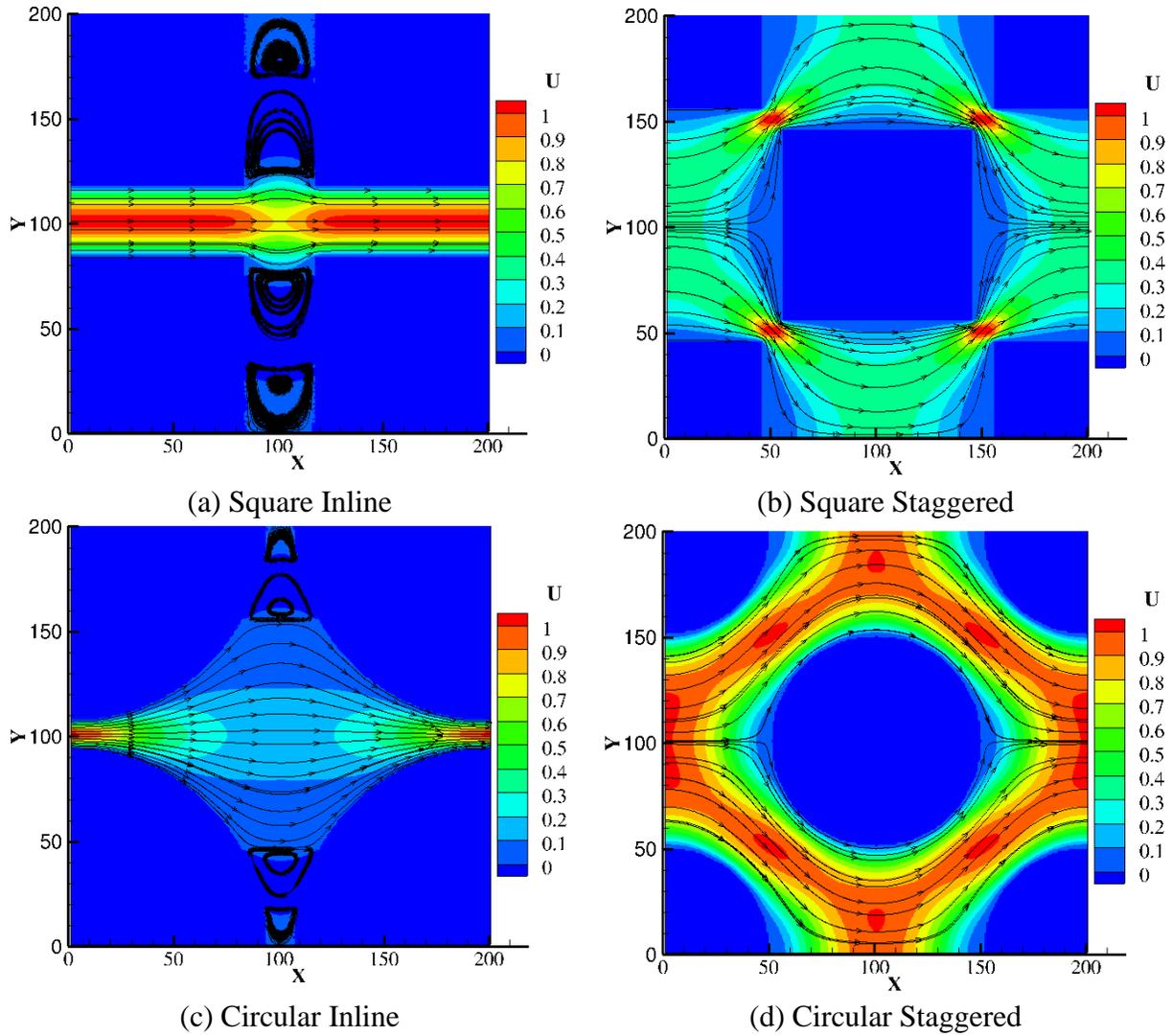

**Figure 6: Streamlines and velocity of Newtonian fluid flow for square and circular cylinders in inline and staggered arragement**

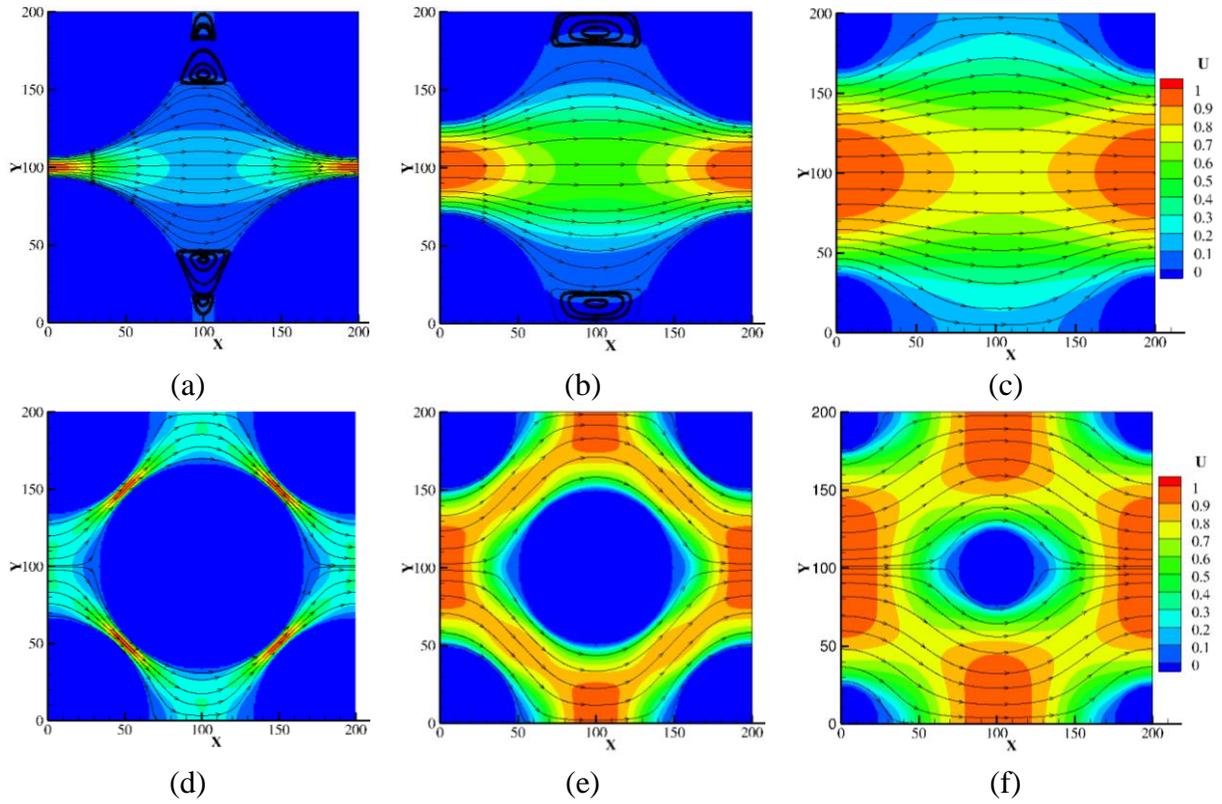

**Figure 7: Streamlines and velocity of non-Newtonian flow for circular cylinders in inline and staggered arragement for porosities 0.3, 0.6 and 0.9**

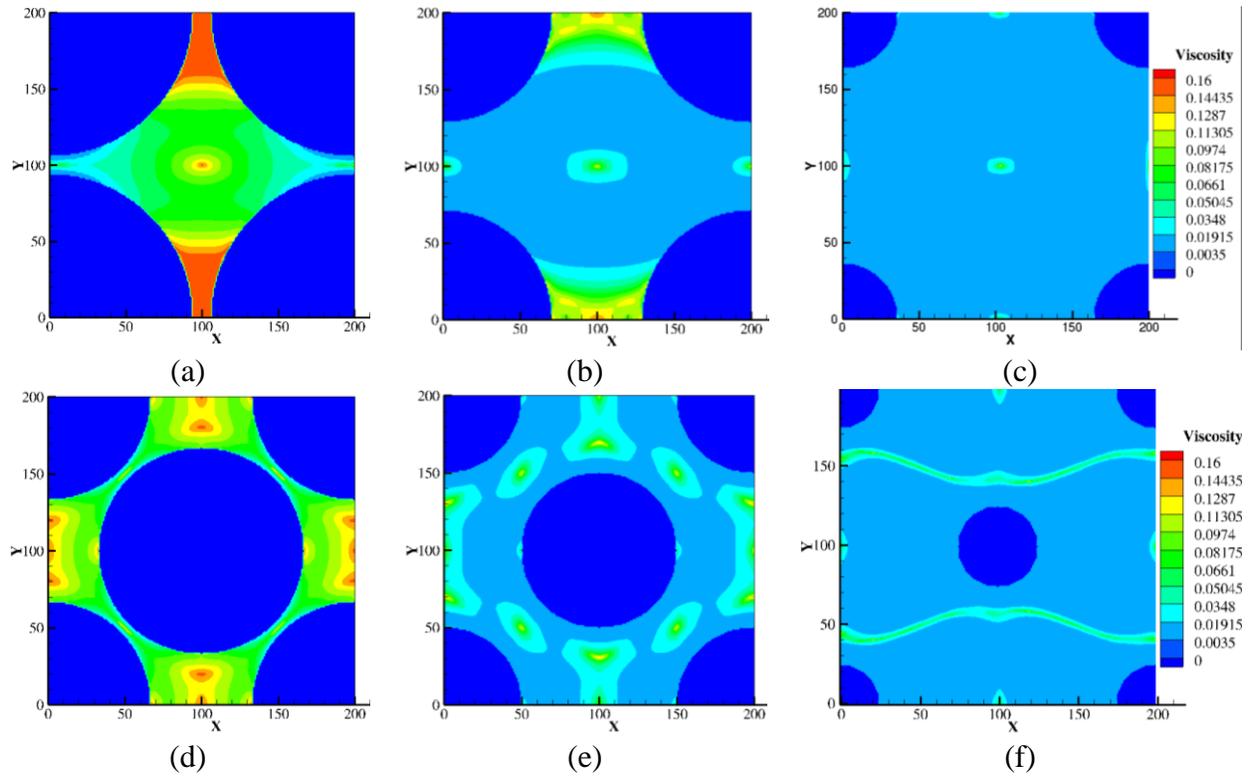

**Figure 8: Distribution of local viscosity of non-Newtonian flow for circular cylinders in inline and staggered arragement for porosities 0.3, 0.6 and 0.9**

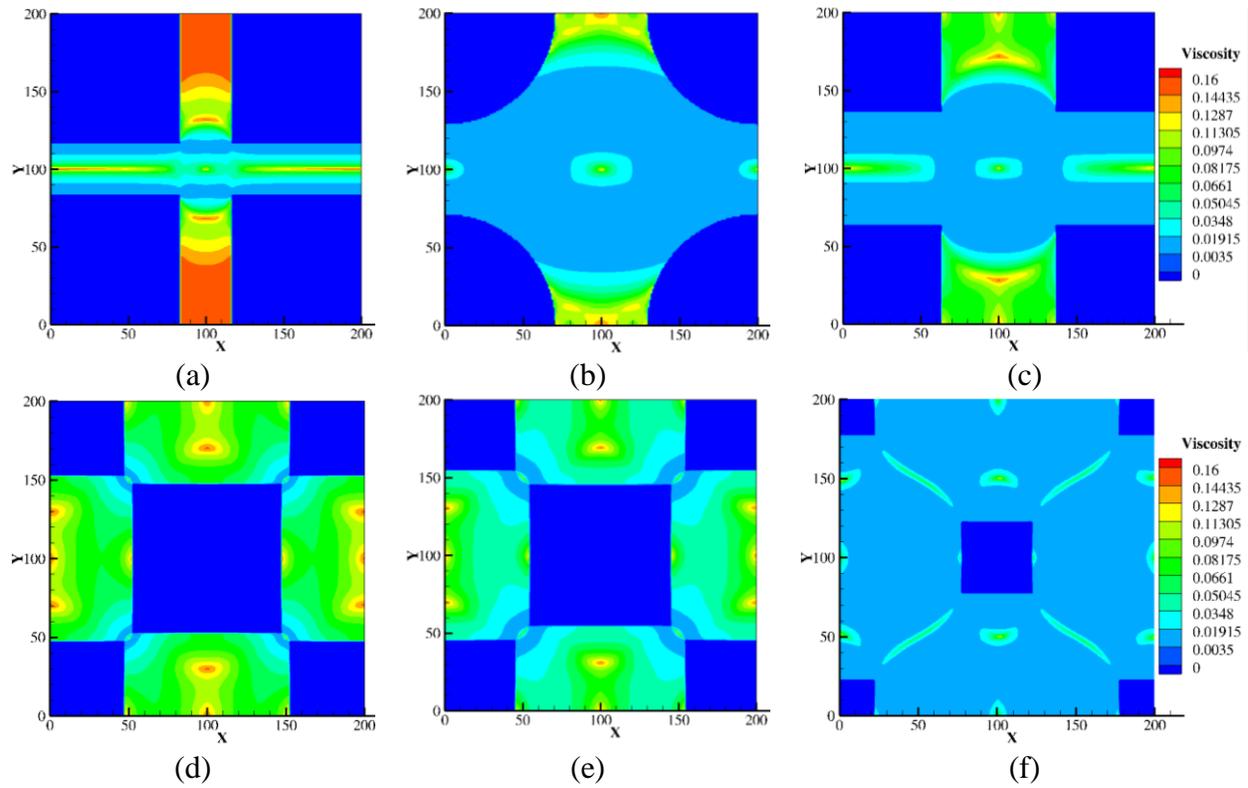

**Figure 9: Distribution of local viscosity of non-Newtonian flow for square cylinders in inline (for porosities 0.3, 0.6, and 0.9) and staggered arragement (for porosities 0.55, 0.6 and 0.9)**

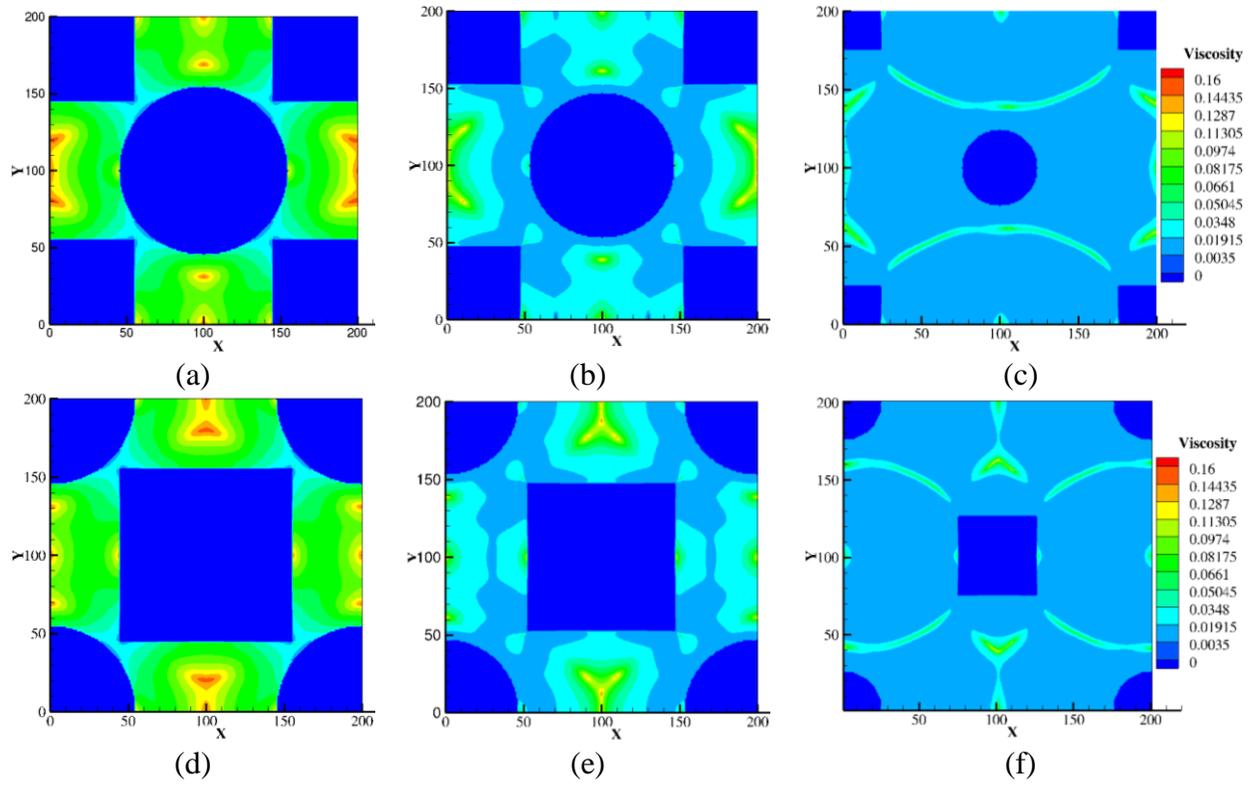

**Figure 10: Distribution of local viscosity of non-Newtonian flow for arragements including both circular and square blocks for porosities 0.45, 0.6 and 0.9**

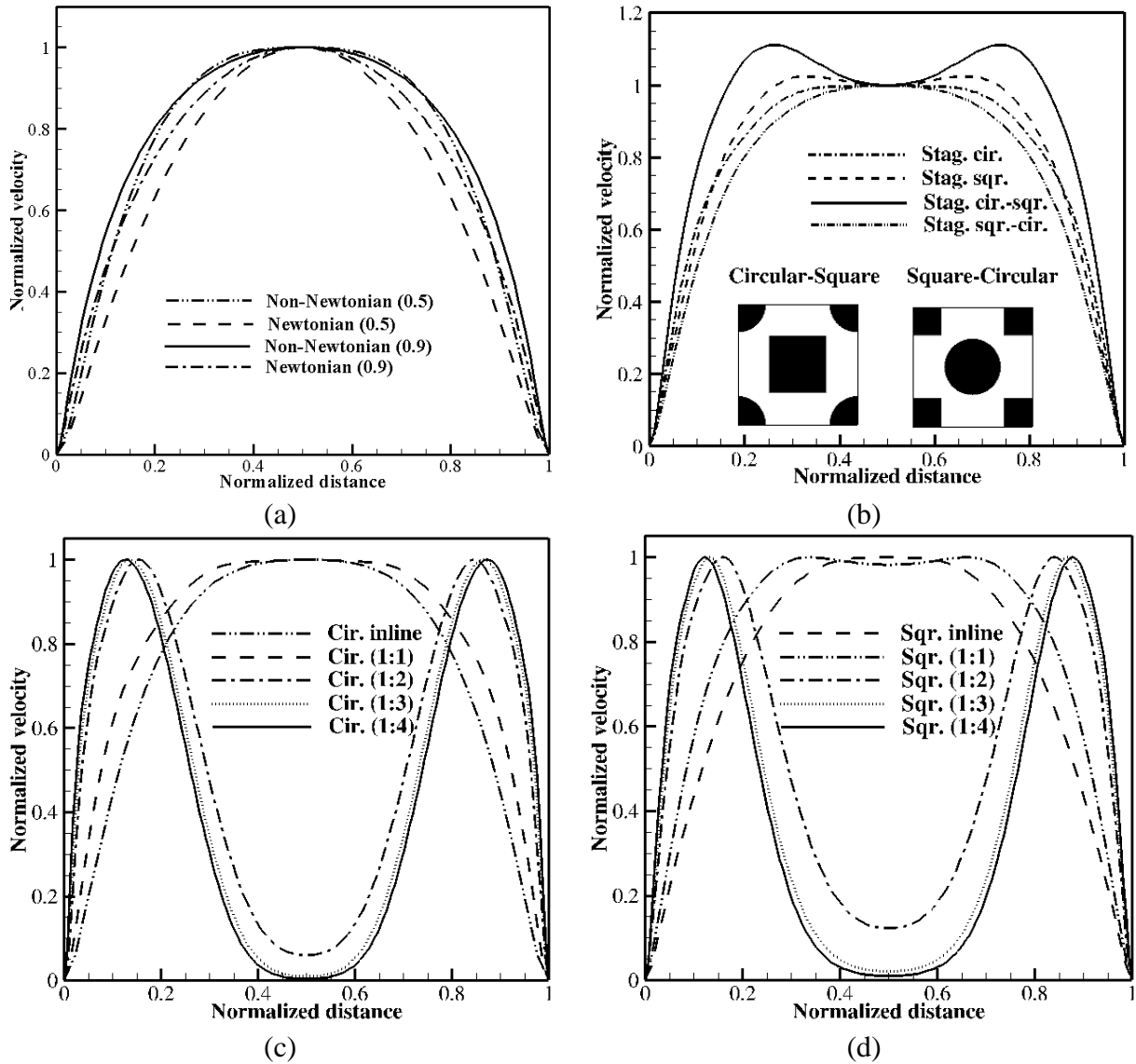

Figure 11: (a) X-direction velocity profile between cylinders along Y for Newtonian and non-Newtonian fluid flow in inline circular cylinders for porosity value of 0.5 and 0.9. (b) Velocity profile for non-Newtonian fluid flow in staggered arrangement for all combinations of cylinders considered. (c) Velocity profile in circular cylinders with different arrangement and inclusion ratio. (d) Velocity profile in square cylinders with different arrangement and inclusion ratio

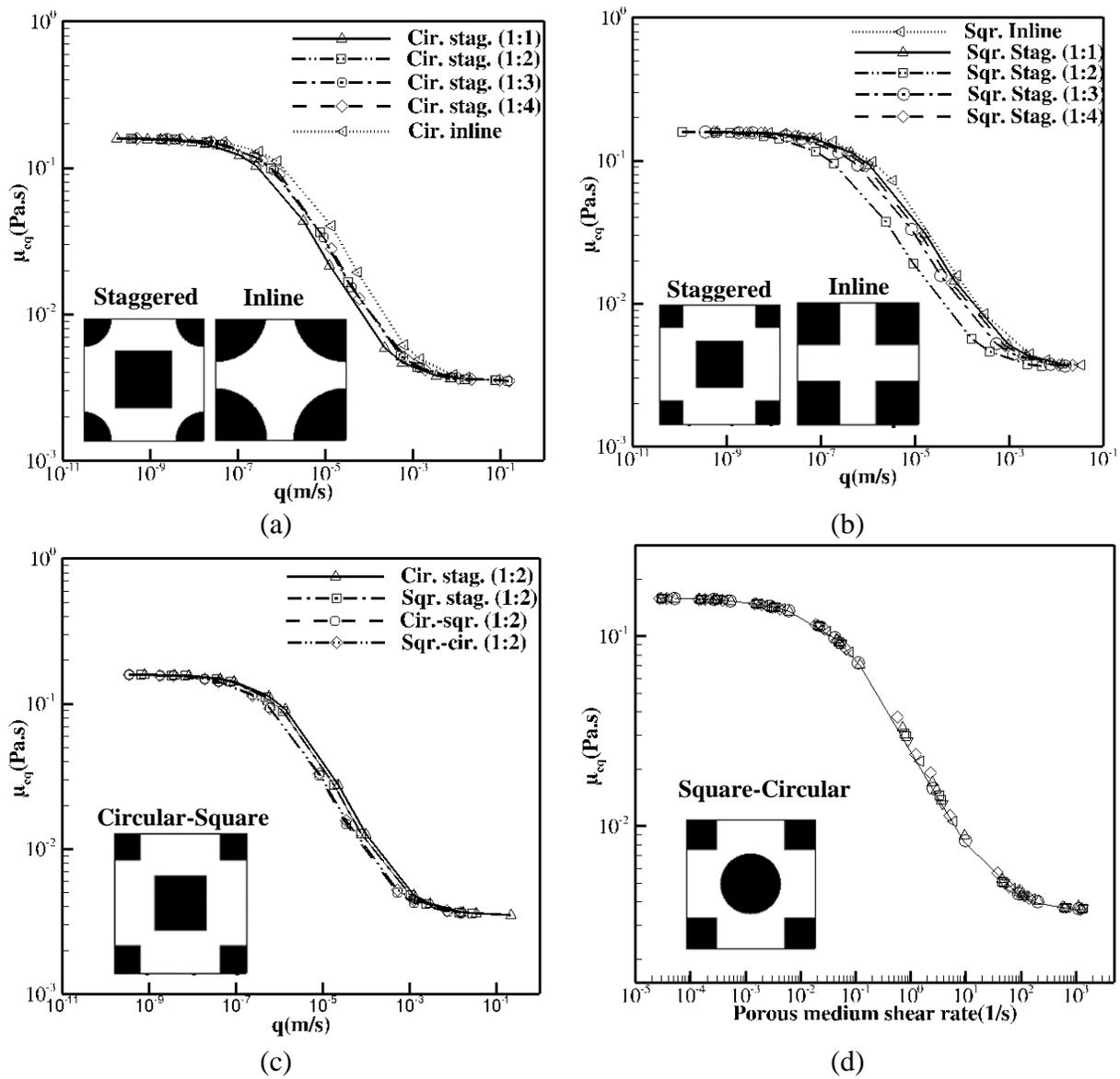

**Figure 12:** Porous medium equivalent viscosity v/s flow rate for (a) circular cylinder with different arrangement, (b) square cylinder with diferent arrangement, (c) different arrangements of cylinders with inclusion ration 1:2. (d) Porous medium viscosity with porous medium shear rate for the circular geometry

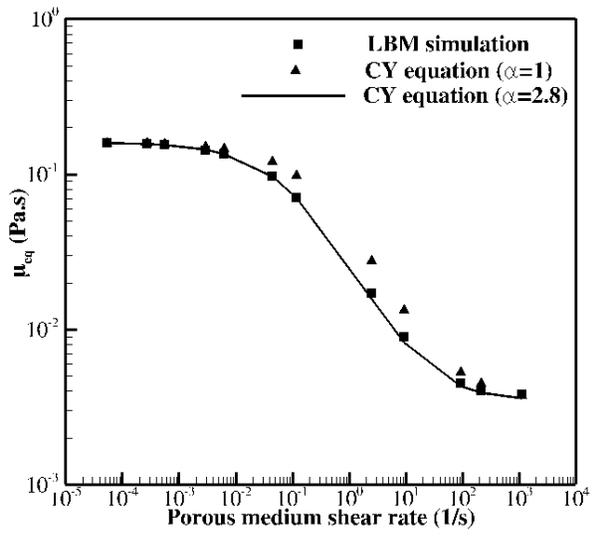 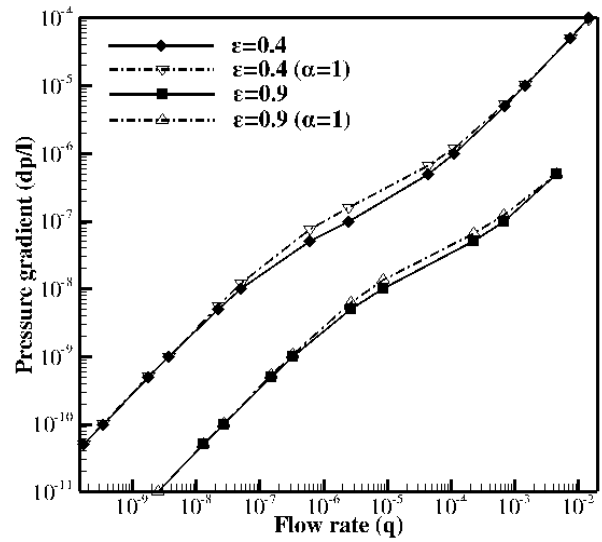

(a) (b)

**Figure 13: Porous medium equivalent viscosity v/s porous medium shear rate is plotted from LBM simulation and Carreau-Yasuda model with and without shift factor**

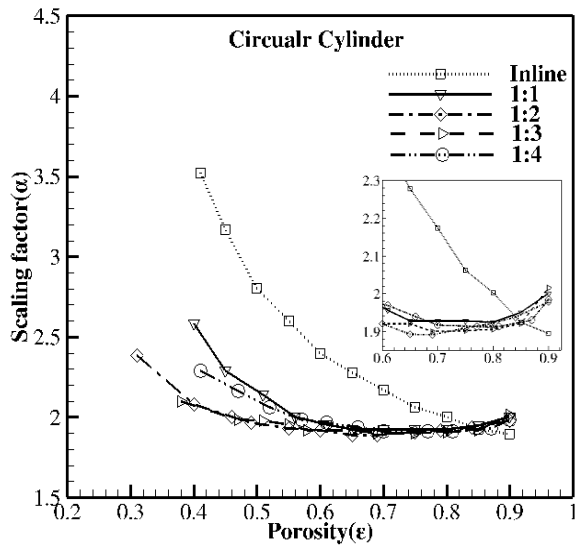
(a)

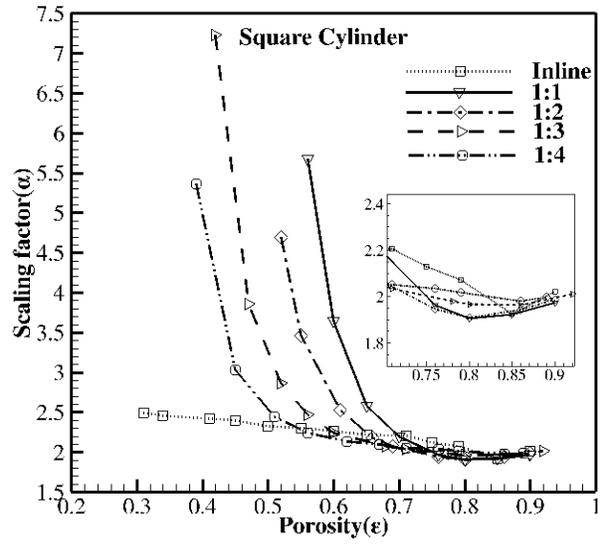
(b)

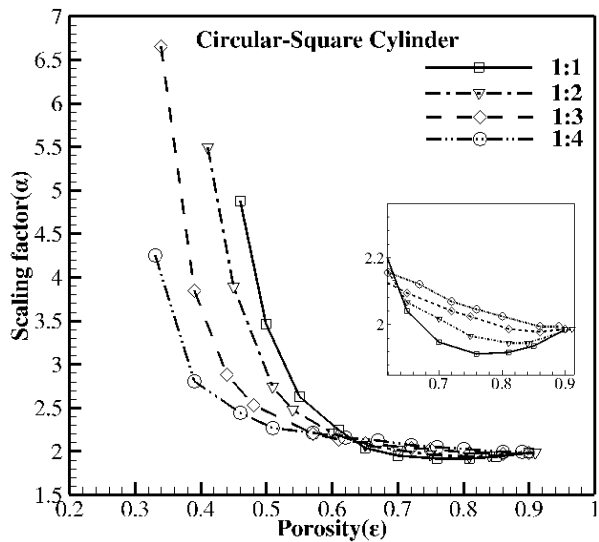
(c)

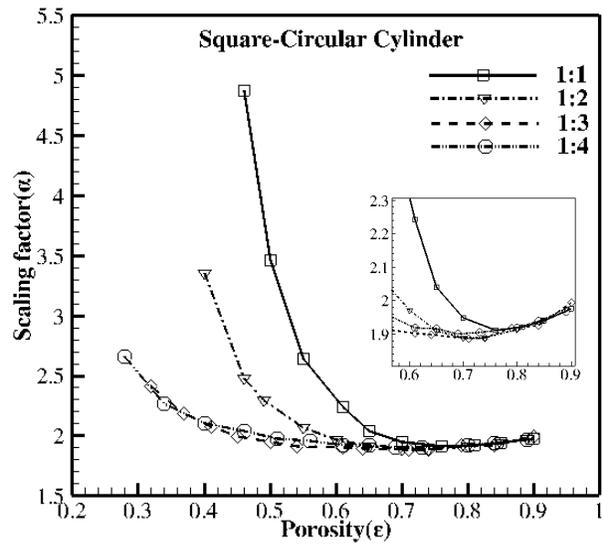
(d)

**Figure 14: Scaling factor is plotted with porosity for the different geometries for various inclusion size ratios. The insets in each figure show the zoomed view of the curves in the high porosity range. The increase in the value of scaling factor for high porosity can be observed from these insets.**

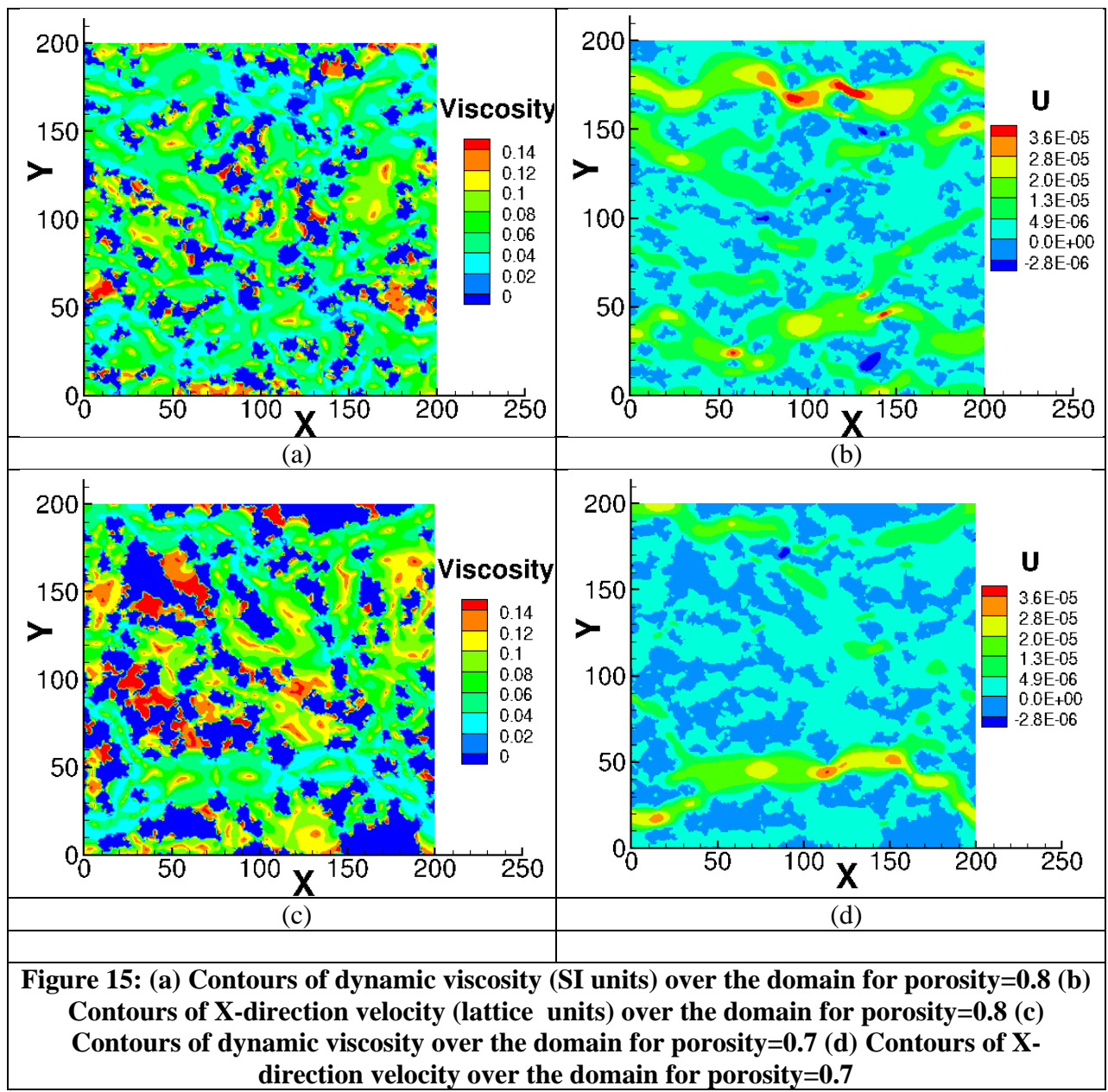

**Figure 15: (a) Contours of dynamic viscosity (SI units) over the domain for porosity=0.8 (b) Contours of X-direction velocity (lattice units) over the domain for porosity=0.8 (c) Contours of dynamic viscosity over the domain for porosity=0.7 (d) Contours of X-direction velocity over the domain for porosity=0.7**

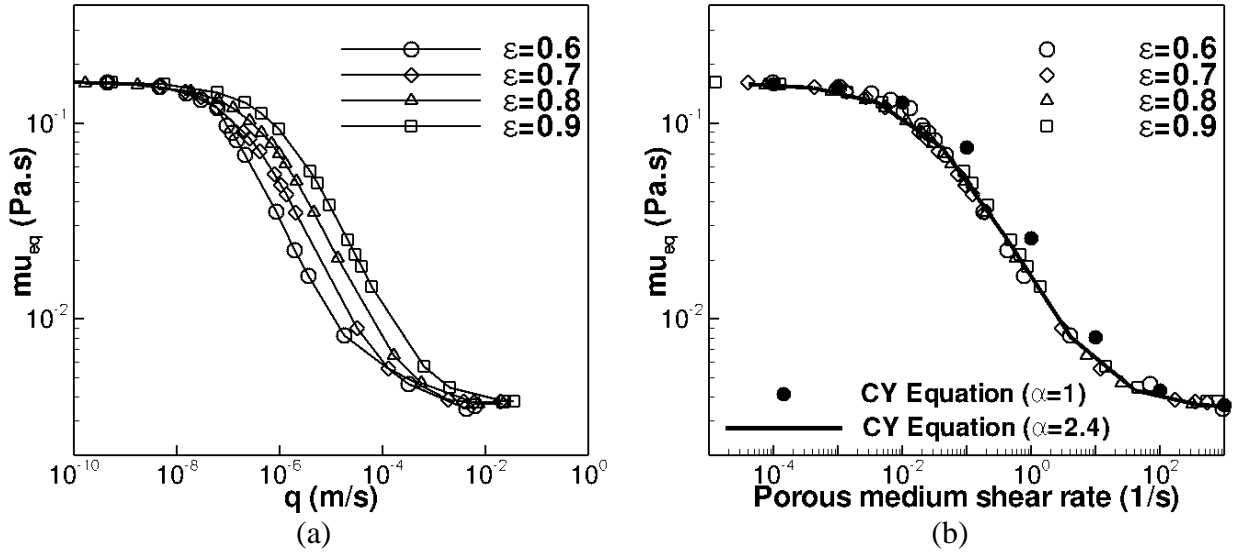

Figure 16: (a) Variation of equivalent porous medium viscosity with flow rates in geometries of various porosities (b) Comparison of equivalent porous medium viscosity values and their variation with porous medium shear rate between simulation and Carreau-Yasuda (CY) equation

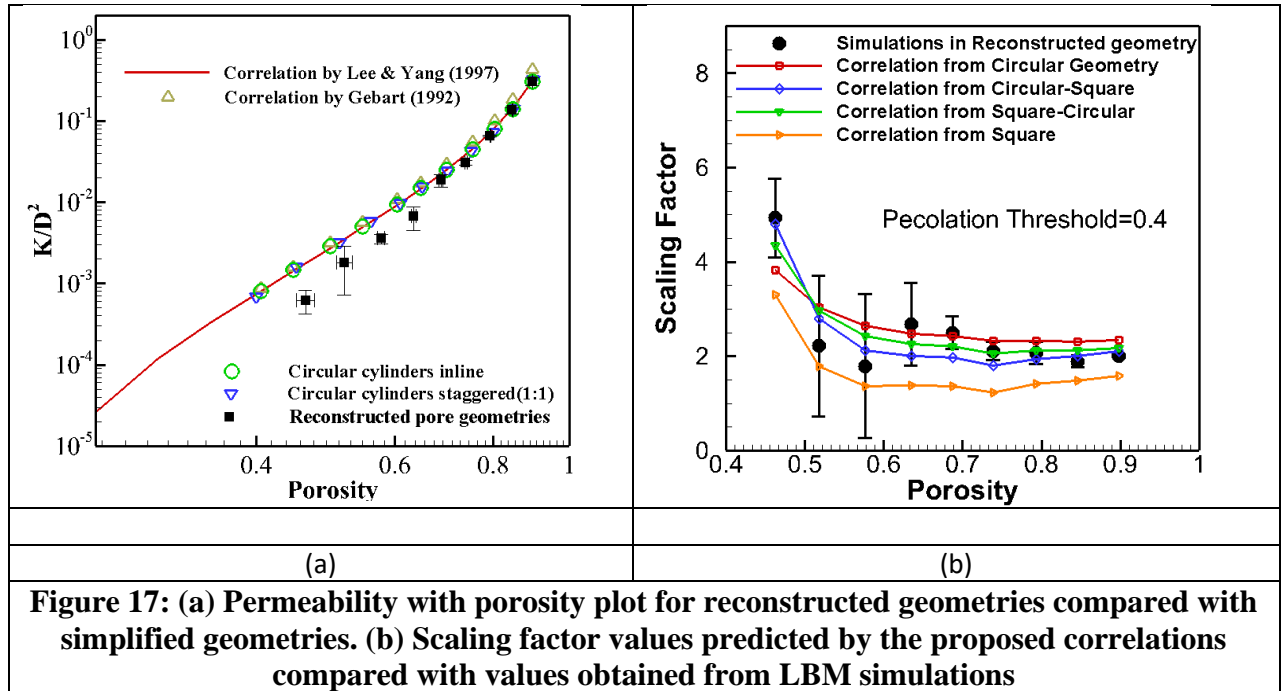

(a) (b)

**Figure 17: (a) Permeability with porosity plot for reconstructed geometries compared with simplified geometries. (b) Scaling factor values predicted by the proposed correlations compared with values obtained from LBM simulations**